\documentclass[11pt, american]{article}%
\usepackage{amssymb}
\usepackage{amsfonts}
\usepackage{amsmath}
\usepackage{lmodern}
\usepackage[nohead]{geometry}
\usepackage[doublespacing]{setspace}
\usepackage[bottom]{footmisc}
\usepackage{booktabs} 
\usepackage{endnotes}
\usepackage{natbib}
\usepackage{graphicx,caption,subfigure}
\usepackage{rotating}
\usepackage{hyperref}
\usepackage[latin1]{inputenc} 
\usepackage[T1]{fontenc} 
\usepackage{pdflscape}
\usepackage{rotating}
\usepackage[flushleft]{threeparttable}
\usepackage{dsfont}
\usepackage{dcolumn}
\newcolumntype{.}{D{.}{.}{4}}
\usepackage{babel}     
\usepackage{ngerman}
\usepackage{array}
\newcounter{rowcount}
\newenvironment{proof}[1][Proof]{\noindent\textbf{#1.} }{\ \rule{0.5em}{0.5em}}
\makeatletter
\def\@biblabel#1{\hspace*{-\labelsep}}
\makeatother
\hypersetup{colorlinks, linkcolor=blue, citecolor=blue, urlcolor=blue}

\DeclareMathOperator{\tr}{tr}
\DeclareMathOperator{\Vect}{Vec}

\usepackage{amsthm}
\theoremstyle{plain}
\newtheorem{assumption}{Assumption}

\newtheorem{lemma}{Lemma}
\theoremstyle{definition}
\newtheorem{proofoflemma}{Proof of Lemma}

\newtheorem{theorem}{Theorem}
\theoremstyle{definition}
\newtheorem{proofof}{Proof of Theorem}

\theoremstyle{definition}

\theoremstyle{definition}

\newtheorem{remark}{Remark}

\begin{document}

\title{Detecting Multiple Structural Breaks\\
in Systems of Linear Regression Equations\\
with Integrated and Stationary Regressors}
\author{Karsten Schweikert\thanks{Address: University of Hohenheim, Core Facility Hohenheim \& Institute of Economics, Schloss Hohenheim 1 C, 70593 Stuttgart, Germany, e-mail: \textit{karsten.schweikert@uni-hohenheim.de}}\medskip\\{\normalsize University of Hohenheim} \date{\normalsize [Latest update: \today]}}
\maketitle

\sloppy

\singlespacing

\begin{abstract}

In this paper, we propose a two-step procedure based on the group LASSO estimator in combination with a backward elimination algorithm to detect multiple structural breaks in linear regressions with multivariate responses. Applying the two-step estimator, we jointly detect the number and location of structural breaks, and provide consistent estimates of the coefficients. Our framework is flexible enough to allow for a mix of integrated and stationary regressors, as well as deterministic terms. Using simulation experiments, we show that the proposed two-step estimator performs competitively against the likelihood-based approach \citep{QuPerron2007, LiPerron2017, OkaPerron2018} in finite samples. However, the two-step estimator is computationally much more efficient. An economic application to the identification of structural breaks in the term structure of interest rates illustrates this methodology.

\strut

\noindent \textbf{Keywords:} LASSO, shrinkage, model selection, cointegration, multivariate \\
\noindent \textbf{JEL Classification:} C32, C52 \\
\noindent \textbf{MSC Classification:} 62E20, 62M10, 91B84

\end{abstract}

\thispagestyle{empty}

\pagebreak
\onehalfspacing

\section{Introduction}\label{sec:intro}
\noindent
Accounting for structural breaks is crucial in time series analysis, particularly in settings involving long spans of data, where the models are more likely to be affected by multiple structural breaks. More specifically, we focus on systems of equations with a mix of integrated and stationary regressors. Thus far, the literature on structural breaks has provided only few methods applicable to linear regressions with multiple equations and integrated regressors \citep[see, for example,][]{BaiLumsdaineStock1998, LiPerron2017, OkaPerron2018}. Without prior knowledge about the structural breaks, methods are needed that precisely determine the number of structural breaks, their timing, and simultaneously estimate the model's coefficients.  

Considering the increasingly larger sample sizes in empirical studies, \cite{MacKinnon2023} cautions against the use of algorithms that are $O(T^{\eta})$ for $\eta \gg 1$ which become impractical for sufficiently large $T$. The currently available methods to solve change-point problems in models that we consider in this paper consist of likelihood-based approaches using dynamic programming techniques \citep{BaiPerron1998, BaiPerron2003} and are characterized by computational costs quadratic in the number of observations. While the approaches are generally very precise, they are not computationally efficient in situations where $T$ is large. Consequently, we propose to lower the computational burden and design a a structural break detection algorithm that is feasible for large $T$ systems.


We consider a penalized regression approach based on the group LASSO estimator to account for multiple structural breaks in such systems which, to the best of our knowledge, has not been explored in the literature yet. Although estimators based on the penalized regression principle have become popular in the context of change-point problems, few prior studies apply them to linear regressions with integrated regressors \citep{SchmidtSchweikert2019, Schweikert2021} or linear regressions with multivariate responses \citep{Gao2019, SafikhaniShojaie2020}. While existing approaches follow a specific-to-general principle utilizing a likelihood-based approach to sequentially increase the number of breaks in a model \citep{BaiLumsdaineStock1998, QuPerron2007, LiPerron2017, OkaPerron2018}, we take a general-to-specific approach shrinking down the number of breakpoint candidates to find the best fitting model. While the likelihood-based approach employs dynamic programming techniques and is computationally efficient in rather short samples with (possibly) many structural breaks, the proposed model selection approach is particularly useful for long samples with a moderate number of structural breaks. It can be shown that it has computational costs linear in the number of observations. Therefore, it is a well-suited solution to account for structural breaks in the long-run relationships between trending variables.

In this paper, we extend the two-step estimator proposed in \cite{ChanYauZhang2014} for univariate structural break autoregressive (SBAR) processes. To do so, we modify the group LARS algorithm specifically tailored for univariate change-point problems and extend it to cover multivariate systems. Moreover, we generalize the model specification and allow for a mix of stationary and integrated regressors as well as deterministic trends. Consequently, our approach is flexible enough to model structural breaks in several special cases like, for example, seemingly unrelated regression (SUR) models and dynamically augmented cointegrating regressions.\footnote{While the model structure, in principle, includes the possibility to consider piece-wise stationary VAR models, our technical analysis relies on Assumption \ref{as:2} stated below which is not compatible with VAR models. We refer to \cite{SafikhaniShojaie2020} who use a slightly different penalty to cover a high-dimensional version of this case.}


The idea to perceive the change-point problem in linear regressions as a model selection problem has spawned a diverse literature \citep[see, for example,][]{Harchaoui2010, BleakleyVert2011, ChanYauZhang2014, SafikhaniShojaie2020, Schweikert2021}. In principle, it is possible to shift and turn the regression hyperplane at every point in time using appropriate indicator variables. Finding the true structural breaks corresponds to selecting relevant indicators and eliminating irrelevant indicators thereby optimizing the fit under sparsity. This leads to a high-dimensional penalized regression model with the total number of parameters of the model close to the number of observations. LASSO-type estimators, introduced by \cite{Tibshirani1996}, have attractive properties in high-dimensional settings with a sparse model structure. Their objective function includes a penalty for nonzero parameters and a tuning parameter controls the sparsity of the selected model. However, quite restrictive regularity conditions about the design matrix (restricted eigenvalue condition \citep{BickelRitovTsybakov2009} or strong irrepresentable condition \citep{ZhaoYu2006}) need to be imposed to ensure simultaneous variable selection and parameter estimation consistency. Unfortunately, these conditions are usually violated in change-point settings, where adjacent columns of the design matrix differ only by one entry and the design matrix is highly collinear if the sample size grows large. Consequently, the conventional LASSO-type estimators need to be improved to both estimate and select the true model consistently.
  
\cite{Harchaoui2010} are among the first to use penalized regression methods to detect structural breaks. They focus on a piecewise constant white noise process and detect structural breaks using a total variation penalty. \cite{BleakleyVert2011} use the group fused LASSO for detection of piecewise constant signals and \cite{ChanYauZhang2014} develop the aforementioned two-step method. 
\cite{Ciuperca2014}, \cite{JinWuShi2016} and \cite{QianSu2016b} consider LASSO-type estimators for the detection of multiple structural breaks in linear regressions. \cite{BehrendtSchweikert2020} propose an alternative strategy to eliminate superfluous breakpoints identified by the group LASSO estimator. They suggest a second step adaptive group LASSO which performs comparably to the backward elimination algorithm suggested in \cite{ChanYauZhang2014}. \cite{SchmidtSchweikert2019} consider cointegration tests in the presence of structural breaks in the long-run relationship and estimate those breaks with an adaptive LASSO estimator. \cite{Schweikert2021} uses the adaptive group LASSO estimator to estimate structural breaks in single-equation cointegrating regressions. The estimator developed in this paper can be understood as an extension of the \cite{Schweikert2021} approach to multiple equation cointegrated systems, so that structural change of more than one equilibrium relationship can be modelled.

Work on multiple structural change models in the context of a system of multivariate equations is relatively scarce. \cite{Quintos1995, Quintos1997} considers a general time-varying structure for the reduced-rank matrix of a vector error correction model (VECM) so that both the cointegrating vector and the adjustment dynamics may change over time. Similarly, \cite{Seo1998} develops a test for changing cointegrating vectors and adjustment coefficients at a single unknown breakpoint. \cite{BaiLumsdaineStock1998} concentrate on dating and estimating a single structural break in vector autoregressions (VARs) and multiple equation cointegrating regressions. \cite{QuPerron2007} consider the restricted quasi-maximum likelihood estimation of and inference for multiple structural changes in a system of equations. 
A sequential break test can be used to determine the number of structural breaks.
In related studies, \cite{EoMorley2015}, \cite{LiPerron2017}, and \cite{OkaPerron2018} extend the likelihood-based approach in several directions.\footnote{\cite{EoMorley2015} propose confidence sets for the timing of structural break estimation in multiple equation regression models. \cite{LiPerron2017} introduce the concept of locally ordered breaks. They model structural breaks in systems of equations with a combination of integrated and stationary regressors, dealing with situations where the breaks cannot be separated by a positive fraction of the sample size. \cite{OkaPerron2018} highlight that the estimation of common breaks allows for a more precise detection of break dates in multivariate systems. They develop common break tests for this assumption. A common break is defined as a point in time at which at least one coefficient from each equation is not restricted to be the same across two adjacent segments.}
    
Recently, the model selection approach has been applied by \cite{Gao2019} and \cite{SafikhaniShojaie2020} to estimate change-points in a piece-wise stationary VARs. While the former study estimates the change-points for each equation separately, thereby decomposing the problem into smaller single-equation problems, the latter uses a fused LASSO penalty to deal with high-dimensional VAR systems. 
    

In the following, we provide a rigorous analysis of the statistical properties of the proposed two-step estimator and extensive simulation experiments to analyze its finite sample properties. We conduct our technical analysis under relatively mild assumptions about the error term process. Prior studies employing LASSO-type estimators to detect structural breaks assume (Gaussian) white noise error terms \citep[see, for example,][]{ChanYauZhang2014, Gao2019, SafikhaniShojaie2020}, which can be useful to model (V)AR processes. However, this assumption is too restrictive in (multiple equations) linear regressions with integrated regressors often having serially correlated errors. Naturally, it becomes more difficult to detect structural breaks if the error term process is serially correlated. Under those assumptions, we show that our estimator is able to consistently estimate the number of structural breaks, their timing, and jointly estimates the model's coefficients. 
    
We use simulation experiments to evaluate our new approach against existing approaches like the likelihood-based approach by, inter alia, \cite{QuPerron2007}, \cite{LiPerron2017}, and \cite{OkaPerron2018}. It is shown that the two-step estimator has competitive finite sample properties with a slight reduction in precision, but substantially improved computational efficiency. Reducing the computational burden over the likelihood-based approach is an important advantage when large sample sizes are available and a moderate number of structural breaks is expected as is often the case in empirical applications involving trending regressors. Another advantage is the joint estimation of the number of breaks, their timing, and the model's coefficients. In the likelihood-based framework, the number of breaks has to be determined based on the evaluation of two tests with the usual implications regarding size and power.\footnote{First, a double maximum test is conducted to test whether at least one break is present, then the exact number of breaks is determined testing the hypothesis of $l$ breaks versus the alternative of $l + 1$ breaks. Naturally, the sequential test procedure requires the specification of a nominal significance level $\alpha$ which implies that also for large samples, the number of breaks is overestimated in roughly $\alpha \cdot 100$\% of all cases.}
In contrast, the approach taken in this paper does not rely on statistical testing, instead we determine the number of breaks as the number of nonzero groups estimated by the group LASSO estimator which is then further reduced by a second step backward elimination algorithm. While we also need to ensure that each identified regime has a sufficient number of observations to estimate the coefficient changes, we need a much smaller trimming parameter than commonly applied in the literature \citep[e.g., 15\% of the sample between breakpoints as suggested in][]{QuPerron2007, OkaPerron2018}. Hence, the two-step estimator is able to detect breaks near the boundary much more reliable than the likelihood-based approach. Furthermore, since we detect structural breaks by the Euclidean norm of the group of coefficient changes, the precision of the group LASSO estimator is mostly determined by the total magnitude of each break. This implies that we do not rely on a distinction between common and partial breaks which is important for the properties of hypothesis tests conducted in the likelihood-based approach to determine the number of breaks. In total, both approaches are conceptually very different so that one approach can serve as a valuable robustness check for the model specification chosen by the other approach. 
    
Finally, we apply the two-step estimator to a term structure model of US interest rates to demonstrate its properties in a real world setting. Relying on the reduced computational burden of the proposed estimator, we are able to estimate the term structure model with daily data over a 30 year span and detect four important structural breaks. Our results reveal substantial differences in the parametrization of the resulting five term structure regimes. The estimates of the proportionality coefficient are smaller than one in every regime and the long-run implications of the expectations hypothesis are rejected in three out of five regimes.


The paper is organized as follows. \autoref{sec:method} outlines the proposed model selection procedure to estimate structural breaks in multivariate systems and presents our main technical results. \autoref{sec:sim} is devoted to the Monte Carlo simulation study. \autoref{sec:emp} reports the results of an empirical application of our methodology to the term structure of US interest rates, and \autoref{sec:conc} concludes. Proofs of all theorems in the paper are provided in the Mathematical Appendix.

\section{Methodology}\label{sec:method}
\noindent
Using penalized regression techniques for structural break detection, we aim to divide a set of breakpoint candidates into two groups of active and inactive breakpoints. Our starting points are \cite{ChanYauZhang2014} and \cite{Schweikert2021}, where a two-step procedure is proposed to detect and estimate multiple structural breaks in autoregressive processes and single equation cointegrating regressions, respectively. Here, the model of interest is a multiple equations system of linear regressions with integrated and stationary regressors, $q$ equations, and $T$ time periods. 

\subsection{First step estimator}\label{sec:first}

We consider the following potentially cointegrated system in triangular form
\begin{eqnarray}
\label{eq:nobreak}
Y_t &=& A X_t + \delta t + \mu + B w_t + u_t, \qquad t = 1, 2, \dots,  \\
\notag X_t &=& X_{t-1} + \xi_t, 
\end{eqnarray}
where $Y_t$ is a $q \times 1$ vector of dependent variables, $X_t$ is a $r \times 1$ vector of integrated regressors, $w_t$ is a $s \times 1$ vector of stationary variables, $u_t$ and $\xi_t$ are I(0) error processes. The coefficient matrices $A$ and $B$ have dimension $q \times r$ and $q \times s$, respectively. Throughout, $\Vert \cdot \Vert$ represents either the Euclidean norm for vectors, i.e., $\Vert x \Vert = (\sum_{i=1}^n x_i^2)^{1/2}$ for $x \in \mathbb{R}^n$ or the Frobenius norm for matrices, i.e., $\Vert A \Vert = [\tr (A A')]^{1/2}$ 
for $A \in \mathbb{R}^{m \times n}$.
We study the asymptotic properties of our estimator under the following assumptions about the involved processes:

\begin{assumption}\label{as:1}
(i) $u_t = \sum\limits_{j=0}^{\infty} C_j \epsilon_{t-j} = C(L) \epsilon_t$, $\xi_t = \sum\limits_{j=0}^{\infty} D_j e_{t-j} = D(L) e_t$, $C(1)$ and $D(1)$ are full rank, $\sum\limits_{j=0}^{\infty} j \Vert C_j \Vert < \infty$ and $\sum\limits_{j=0}^{\infty} j \Vert D_j \Vert < \infty$, $(\epsilon_t, e_t)$ are i.i.d.\ with finite $4 + a$ $(a >0)$ moment. $w_t$ is a mean-zero second order stationary process with uniformly bounded $4 + \delta$ moment. The long-run covariance matrix $\Omega_{\xi} = \sum_{j=-\infty}^{\infty} E \xi_t \xi_{t-j}'$ is positive definite. \\
(ii) Further, we require that
\begin{equation*}
\underset{T}{\sup} E \left\vert \frac{1}{T} \sum\limits_{i=1}^{t} X_{l,i} u_i \right\vert^{4+\epsilon} < \infty, \qquad \text{for } 1 \leq l \leq r, \, 1 \leq t \leq T \text{ and some } \epsilon > 0,
\end{equation*}
and
\begin{equation*}
\underset{T}{\sup} E \left\vert \frac{1}{T} \sum\limits_{i=1}^{t} w_{l,i} u_i \right\vert^{4+\epsilon} < \infty, \qquad \text{for } 1 \leq l \leq s, \, 1 \leq t \leq T \text{ and some } \epsilon > 0.
\end{equation*}
\end{assumption}

\begin{assumption}\label{as:2}
The error process $u_t$ is independent of the regressors for all leads and lags.
\end{assumption}

The error term processes are assumed to be linear processes in Assumption \ref{as:1}, satisfying the required conditions to ensure the validity of the functional central limit theorem for partial sum processes constructed from them (see, for example, Theorem 3.4 in \cite{PhillipsSolo1992} and its multivariate extension in \cite{Phillips1995}). Further, $w_t$ is given as a stationary process with a sufficiently well-behaved distribution. Note that these assumptions could be replaced by other sufficient conditions that conform with a strong invariance principle or functional central limit theorem. We assume that $\Omega_{\xi}$ is positive definite which implies that $X_t$ is non-cointegrated. This ensures that the coefficients of $X_t$ have the standard $T$ rate of convergence. Assumption \ref{as:2} is chosen for technical reasons but it is less restrictive than it initially seems considering that $w_t$ might include the leads and lags of changes in $X_t$ (see, for example, \cite{Saikkonen1991}, \cite{PhillipsLoretan1991}, and \cite{StockWatson1993} for a treatment of second-order biased estimators in the context of cointegrating regressions).\footnote{We note that Assumption \ref{as:2} is stronger than the moment conditions given in Assumption \ref{as:1} (ii) and replaces them in the corresponding results (Theorem \ref{th:2} -- Theorem \ref{th:4}). However, Assumption \ref{as:2} is not necessary to proof Theorem \ref{th:1}.}

Assumption \ref{as:1} allows for serially correlated error terms but this conflicts with endogenous regressors like lagged dependent variables. The results presented in the following remain valid for endogenous regressors as long as the errors are not permitted to be serially correlated \citep[][discuss this issue on page 51]{BaiPerron1998}.\footnote{The simulation results presented in Table S6 in Supplementary Material A show that the structural break detection remains consistent for endogenous regressors and white noise errors.}

We follow \cite{LiPerron2017} and \cite{OkaPerron2018} and apply scaling factors in Equation~\eqref{eq:nobreak} so that the order of all regressors is the same.\footnote{Higher order deterministic trends can be included in the model in the same way as long as the corresponding scaling factors are applied.} To simplify notation, we write the system of equations in its stacked form as
\begin{equation}
Y_t = (Z_t' \otimes I) \theta + u_t,
\end{equation}
where $Z_t = (T^{-1/2} X_t', T^{-1}t, 1, w_t')'$ and $\theta = \Vect(A, \delta, \mu, B)$ is a $d = q(r + 2 + s)$ column vector, concatenating the coefficients for each regressor over all equations. The operator $\Vect(\cdot)$ stacks the rows of a matrix into a column vector, $\otimes$ denotes the Kronecker product and $I$ is a $q \times q$ identity matrix. While model~\eqref{eq:nobreak} allows for very flexible specifications and covers several special cases (e.g., SUR models for $A = 0$ and $\delta = 0$, VAR(s) models for $A = 0$ and $\delta = 0$ and $w_t = (Y_{t-1}, \dots, Y_{t-s})'$, or pure cointegrating regressions for $\delta = \mu = 0$ and $B = 0$), our main focus is on a full model specification with a mix of integrated and stationary regressors. For example, if $w_t$ contains the leads and lags of changes in $X_t$, we estimate structural breaks in a dynamically augmented cointegrating regression with multivariate responses. Assumption \ref{as:2} is violated in VAR models which invalidates our technical analysis for this specification. Hence, we refer to \cite{SafikhaniShojaie2020} for the appropriate assumptions to show that LASSO-type estimators can be used to estimate piece-wise stationary VAR models.  
If it is known to the researcher that some of the coefficients do not change, it is possible to introduce a selection matrix $S$ to consider only partial structural breaks in the sense that some coefficients are constant over the entire sampling period. This allows the researcher to estimate the respective coefficients with full efficiency. The selection matrix contains elements that are either 0 or 1 and hence, specifies which regressors appear in each equation.\footnote{Note that $S'S$ is idempotent with non zero elements only on the diagonal. The rank of $S$ is equal to the number of coefficients that are allowed to change.} For example, we can use the selection matrix to focus only on breaks in the matrix $A$, i.e., detecting breaks in the long-run coefficients of a cointegrating regression but leave the matrix $B$ constant over the sample period. 

We assume that the system includes $m_0$ true structural breaks. Multiple (partial) structural breaks in the regression coefficients can be expressed using the following model 
\begin{equation}
Y_t = (Z_t' \otimes I) \theta_{t_0} + \sum\limits_{k=1}^{m} d(t_k) (Z_t' \otimes I) S' S \theta_{t_k} + u_t,
\label{eq:break}
\end{equation} 
where $d(t_k) = 0$ for $t \leq t_k$ and $d(t_k) = 1$ for $t \geq t_k$. The total number of potential structural breaks or breakpoint candidates in this model is denoted by $m$, $\theta_{t_0}$ is the baseline coefficient vector, and $\theta_{t_k}$, $k = 1, \dots, m$ are regime-dependent changes in the regression coefficients. In situations where $m > m_0$, our structural break model in Equation~\eqref{eq:break} considers more structural breaks than necessary which implies that the true coefficient vector does not change at some $t_k$, i.e., $\theta^0_{t_k} = 0$. For the breakpoints or break dates $t_k$, it holds by general convention that $1 = t_0 < t_1 < \dots < t_m < t_{m+1} = T + 1$. The relative timing of breakpoints is denoted by $\tau_k = t_k / T, k \in \lbrace 1, \dots m \rbrace$. We assume that there is a change in at least one coefficient matrix at each true structural break, so that $\Vert S \theta^0_{t_k} \Vert \neq 0$. To simplify the notation, we assume that all coefficients change at all breakpoints for the remainder of the paper.


In case of unknown number and timing of structural breaks, each point in time has to be considered as a potential breakpoint. Therefore, it is helpful to estimate the model in Equation~\eqref{eq:break} with $m = T$ under the condition that the set $\theta(T) = \lbrace \theta_1, \theta_2, \dots, \theta_T \rbrace$ exhibits a certain sparse nature so that the total number of distinct vectors in the set equals the true number of breaks $m_0$. To use a convenient matrix notation, we define
\begin{equation}
\mathcal{Z} = \begin{pmatrix}
Z_1' & 0 & 0 & \dots & 0 \\
Z_2' & Z_2' & 0 & \dots & 0 \\
Z_3' & Z_3' & Z_3' & \dots & 0 \\
\vdots\\
Z_T' & Z_T' & Z_T' & \dots & Z_T' \\
\end{pmatrix},
\end{equation}
$\mathcal{Y} = (Y_1', \dots, Y_T')'$, $\mathcal{U} = (u_1', \dots, u_T')'$ and $\boldsymbol{\theta}(T) = (\boldsymbol{\theta}_1', \dots, \boldsymbol{\theta}_T')'$. Furthermore, we define $\boldsymbol{Y} = \Vect(\mathcal{Y})$, $\boldsymbol{Z} = I \otimes \mathcal{Z}$, and $\boldsymbol{U} = \Vect(\mathcal{U})$. Now, the system for $m=T$ breakpoint candidates can be rewritten as
\begin{equation}
\boldsymbol{Y} = \boldsymbol{Z} \boldsymbol{\theta}(T) + \boldsymbol{U},
\end{equation} 
where $\boldsymbol{Y} \in \mathbb{R}^{Tq \times 1}$, $\boldsymbol{Z} \in \mathbb{R}^{Tq \times Td}$, $\boldsymbol{\theta}(T) \in \mathbb{R}^{Td \times 1}$, and $\boldsymbol{U} \in \mathbb{R}^{Tq \times 1}$. Note that this model specification reorders the groups so that $\boldsymbol{\theta}(T)$ contains the breakpoint candidates for each equation successively, i.e. $\boldsymbol{\theta}_i = \boldsymbol{K} \theta_i$ with commutation matrix $\boldsymbol{K}$. 
We assume that at least one of the baseline coefficients is nonzero to distinguish between active and inactive breakpoints without making any case-by-case considerations. This implies that the vector of true coefficients $\boldsymbol{\theta}^0(T)$ contains $m_0 + 1$ nonzero groups and $\boldsymbol{\theta}_1 \neq \boldsymbol{0}$. For the remainder of this paper, $\boldsymbol{\theta}_i = \boldsymbol{0}$ means that $\boldsymbol{\theta}_i$ has all entries equaling zero and $\boldsymbol{\theta}_i \neq \boldsymbol{0}$ means that $\boldsymbol{\theta}_i$ has at least one non-zero entry. We define the index sets $\bar{\mathcal{A}} = \lbrace 1 \leq i \leq T: \boldsymbol{\theta}^0_i \neq \boldsymbol{0} \rbrace$ denoting the indices of truly non-zero coefficients (including the baseline coefficient) and $\mathcal{A} = \lbrace i \geq 2: \boldsymbol{\theta}^0_i \neq \boldsymbol{0} \rbrace$ denoting the non-zero parameter changes. The set $\mathcal{A}_T = \lbrace \hat{t}_1, \hat{t}_2, \dots, \hat{t}_{\hat{m}} \rbrace$ denotes the $\hat{m}$ breakpoints estimated in the first step, i.e., indices of those coefficient changes which are estimated to be non-zero. $|\mathcal{A}|$ denotes the cardinality of the set $\mathcal{A}$ and $\mathcal{A}^c$ denotes the complementary set.

We propose to estimate the set of coefficient changes $\boldsymbol{\theta}(T)$ by minimizing the following penalized least squares objective function \citep{YuanLin2006}:
\begin{eqnarray}
Q^*(\boldsymbol{\theta}(T)) &=& \frac{1}{T} \left\Vert \boldsymbol{Y} - \boldsymbol{Z} \boldsymbol{\theta}(T) \right\Vert^2 + \lambda_T \sum\limits_{i=1}^{T} \Vert \boldsymbol{\theta}_i \Vert,
\label{eq:gfl2}
\end{eqnarray}
where $\lambda_T$ is a tuning parameter and $\Vert \cdot \Vert$ denotes the $L_2$-norm. 
Minimizing the objective function in \eqref{eq:gfl2} yields the group LASSO estimator which is denoted by $\hat{\boldsymbol{\theta}}(T)$. Using a group penalty for $\boldsymbol{\theta}_i$, in principle, we assume common breaks across all equations as defined by \cite{QuPerron2007} and \cite{OkaPerron2018}. However, it should be noted that the issue of common or partial breaks is crucial for the specification of hypothesis tests required for the likelihood-based approach to detect the true number of breaks but it is not as important for the group LASSO estimator. Since we detect structural breaks by the Euclidean norm of the group of coefficient changes (vectorizing all coefficient matrices), only the total magnitude of each break enters our objective function. Consequently, prior knowledge about partial breaks would not improve our sensitivity detecting those breaks as much as it would in the likelihood-based approach.

Using these definitions, we frame the detection of structural breaks as a model selection problem and are able to use efficient algorithms from this strand of the literature \citep{HuangBrehenyMa2012, ChanYauZhang2014, YauHui2017} to eliminate irrelevant breakpoint candidates. Depending on the value of $\lambda_T$, a sparse solution is obtained so that the number of nonzero groups corresponds to the number of estimated breaks and the coefficient changes at each breakpoint are contained within the nonzero groups.\footnote{Practical guidance for the choice of $\lambda_T$ is given in \autoref{sec:sim}.} In the next subsection, we investigate the asymptotic properties of the first step estimator in this setting.

\subsection{Asymptotic properties of the first step estimator}\label{sec:asymptotics}

We show in the following that the group LASSO estimator for structural breaks in a system of linear regression equations is consistent in terms of prediction error but inherits the same problems, namely estimation inefficiency and model selection inconsistency, as shown for univariate AR models \citep{ChanYauZhang2014}, single-equation cointegrating regressions \citep{Schweikert2021}, and piecewise-stationary VAR models \citep{Gao2019}. As discussed in \cite{ChanYauZhang2014}, any two adjacent columns of the matrix $\mathcal{Z}$ only differ by one entry. Consequently, the restricted eigenvalue condition \citep{BickelRitovTsybakov2009} does not hold in our setting and we cannot establish our consistency proofs based on this assumption. 

Further assumptions about the timing of true breakpoints ($\tau^0_k$, $k = 1, \dots, m_0$) and the magnitude of coefficient changes have to be stated to continue our analysis. 

\begin{assumption}\label{as:5}
(i) The break magnitudes are bounded to satisfy \\
$m_{\theta} = \min_{1 \leq j \leq m_0 + 1} \Vert \boldsymbol{\theta}^0_{t^0_{j-1}} \Vert \geq \nu > 0$ and $M_{\theta} = \max_{1 \leq j \leq m_0 + 1} \Vert \boldsymbol{\theta}^0_{t^0_{j-1}} \Vert \leq \mathcal{V} < \infty$. \\
(ii) $\min_{1 \leq j \leq m_0+1} \vert t^0_j - t^0_{j-1} \vert / T \gamma_T \to \infty$ for some $\gamma_T \to 0$ and $\gamma_T/\lambda_T \to \infty$ as $T \to \infty$.
\end{assumption}

The first inequality of Assumption \ref{as:5}(i) is a necessary condition to ensure that a structural break occurs at $t_j^0$ and the second part excludes the possibility of infinitely large parameter changes.\footnote{Our definitions in Assumption \ref{as:5}(i) include the baseline coefficients which can, in principle, be relaxed but simplifies the technical analysis.} 
We do not consider breaks with local-to-zero behaviour in this setting (see \cite{BaiLumsdaineStock1998} for assumptions used in this context). This assumption is not believed to be restrictive for the intended empirical applications. In the case of a full specification with integrated regressors, applied researchers aim to estimate structural long-run relationships that are often needed for their follow-up analysis, e.g., in a cointegrated VAR as in \cite{Hansen2003}. Essentially, they need optimal in-sample forecasts in terms of mean squared error of the cointegrating regressions under structural instability to consistently estimate deviations from the long-run equilibrium. \cite{BootPick2019} show that in-sample forecasts are largely unaffected by local-to-zero breaks. Assumption \ref{as:5}(ii) requires that the length of the regimes between breaks increases with the sample size albeit slower than $T$. If $\gamma_T$ is chosen with a slow enough rate, which depends on the tuning parameter sequence $\lambda_T$, it allows us to consistently detect and estimate the true break fractions.

The first result for the first step estimator shows that it is consistent in terms of prediction error if the tuning parameter $\lambda_T$ grows at the right rate.

\begin{theorem}\label{th:1}
Under Assumption \ref{as:1} and Assumption \ref{as:5}, if $\lambda_T = 2 d c_0 (\log T / T)^{1/4}$ for some $c_0 > 0$, then there exists some $C > 0$ such that with probability greater than $1 - \frac{C}{c_0^4 \log T}$,
\begin{equation*}
\frac{1}{T} \Vert \boldsymbol{Z} \left( \hat{\boldsymbol{\theta}}(T) - \boldsymbol{\theta}^0(T) \right) \Vert^2 \leq 4 d c_0 \left( \frac{\log T}{T} \right)^{\frac{1}{4}} (m_0 + 1) M_{\theta}.
\end{equation*}
\end{theorem}

\begin{remark}
The corresponding convergence rate for univariate piecewise stationary autoregressive processes given in \cite{ChanYauZhang2014} is $\sqrt{\log T / T}$. However, they assume $\beta$-mixing stationary processes with a white noise error term so that the necessary tail bounds can be derived. Instead, Assumption \ref{as:1} in this paper only requires that the error terms of the system follow stationary processes with some additional moment conditions, i.e., allows for serially correlated errors. The convergence rate given in \cite{SafikhaniShojaie2020} is not directly comparable because the VAR system is assumed to be high-dimensional with the number of equations increasing with $T$ and a different penalty is used to handle the growing number of time series. Expectedly, \cite{Gao2019} find the same rate as \cite{ChanYauZhang2014}.
\end{remark}

The next result shows that the number of estimated breaks is at least as large as the true number of breaks. Furthermore, the location of the breakpoints can be estimated within an $T\gamma_T$-neighborhood of their true location. To state the theorem, we have to define the Hausdorff distance between the set of estimated breakpoints and the set of true breakpoints. We follow \cite{Boysen2009} and define $d_H(A, B) = \underset{b \in B}{\max} \, \underset{a \in A}{\min} |b - a|$ with $d_H(A, \emptyset) = d_H(\emptyset, B) = 1$, where $\emptyset$ is the empty set.

\begin{theorem}\label{th:2}
If Assumptions \ref{as:1} - \ref{as:5} hold, then as $T \to \infty$
\begin{equation*}
P \left( |\mathcal{A}_T| \geq m_0 \right) \to 1,
\end{equation*}
and 
\begin{equation*}
P \left( d_H(\mathcal{A}_T, \mathcal{A}) \leq T\gamma_T \right) \to 1.
\end{equation*}
\end{theorem}

\begin{remark}
When $m_0$ is known, we can show that the breakpoints are estimated with the same convergence rate (Lemma \ref{lemma:5} in the Mathematical Appendix). Setting $\gamma_T = \log T/T$ and $\lambda_T = O(\log T / (T \log \log T))$, the convergence rate is identical to the one found by \cite{ChanYauZhang2014}, assuming Gaussian white noise errors. 
Note that the tuning parameter can be set differently than in Theorem \ref{th:1} as the consistency in prediction error is not important for the proof. For the optimal convergence rate, the tuning parameter must be set conforming to a rate that has a higher order than the one in Theorem \ref{th:1} and thus penalizes the coefficient changes more strongly. Finally, \cite{SafikhaniShojaie2020} show that the rate could be as low as $(\log \log T \log q)/T$ for high-dimensional piecewise stationary VAR processes with Gaussian white noise error terms, where $q$ denotes the number of variables in the VAR system and $q > T$.
\end{remark}

\begin{remark}
The second part of Theorem \ref{th:2} implies that the Hausdorff distance from the set of estimated breakpoints to the true breakpoints diverges slower than the sample size. Consequently, the Hausdorff distance of the relative location to their true location of the breakpoints converges to zero at rate $\log T / T$ and therefore is only slightly above the optimal rate $1/T$ for estimators of break fractions in regression models \citep{Bai1997, Bai2000}. This provides us with a consistency result for the estimated break fractions in the first step and gives us justification to consider multiple structural breaks at once, since the Hausdorff distance evaluates the joint location of all breakpoints.
\end{remark}

\subsection{Second step estimator}\label{sec:asymptotics_second}

To obtain a consistent estimator for the number of breaks, their timing and coefficient changes, we need to design a second step refinement reducing the number of superfluous breaks. Immediate candidates are using a backward elimination algorithm (BEA) optimizing some information criterion \citep{ChanYauZhang2014, Gao2019, SafikhaniShojaie2020} or applying the adaptive group LASSO estimator as a second step using the group LASSO estimates as weights \citep{BehrendtSchweikert2020, Schweikert2021}. In the following, we outline the former approach and discuss the latter approach in the Supplementary Material B because the BEA produces more accurate results in our simulation experiments.\footnote{Supplementary Material B can be found here: \url{https://karstenschweikert.github.io/mequ_ci/mequ_ci_suppB.pdf}}

According to Theorem \ref{th:2}, the group LASSO estimator slightly overselects breaks under the right tuning. To distinguish between active and non-active breakpoints in the set $\mathcal{A}_T$, we employ an information criterion for the second step which consists of a goodness-of-fit measure, here the sum of squared residuals, and a penalty term as a function of the number of breaks. We define $\widehat{\widehat{\boldsymbol{\theta}}}_j$, $1 \leq j \leq m$ as the least squares estimator of $\boldsymbol{\theta}^0_j$, based on breakpoints estimated in the first step. Further, we define the sum of squared residuals over all $q$ equations as 
\begin{equation}
    S_T(t_1, \dots, t_m) = \sum\limits_{j=1}^{m + 1} \sum\limits_{t = t_{j-1}}^{t_j - 1} \Vert Y_t - \bar{Z}_t \sum\limits_{s=1}^{j} \boldsymbol{K}' \widehat{\widehat{\boldsymbol{\theta}}}_s \Vert^2,
\end{equation}
where $\bar{Z}_t = (Z_t' \otimes I)$. For $m$ and the breakpoints $\boldsymbol{t} = (t_1, \dots, t_m)$, we can define the information criterion (IC)
\begin{equation}
    IC(m, \boldsymbol{t}) = S_T(t_1, \dots, t_m) + m \omega_T,
    \label{eq:IC}
\end{equation}
where $\omega_T$ is the penalty term that is further characterized below in Theorem \ref{th:3}. We estimate the number of breaks and the timing by solving
\begin{equation}
    (\widehat{\widehat{m}}, \widehat{\widehat{\boldsymbol{t}}}) = \arg \min\limits_{\substack{m \in \lbrace 1, \dots, |\mathcal{A}_T| \rbrace \\ \boldsymbol{t} = (t_1, \dots, t_m) \subset \mathcal{A}_T}} IC(m, \boldsymbol{t}).
\end{equation}
If the maximum number of breaks in the first step algorithm is chosen to be small, the evaluation of the information criterion for each combination of breakpoints can be achieved easily. The following result shows that minimizing the IC gives us a consistent estimator for $m_0$ and $\mathcal{A}$.

\begin{theorem}\label{th:3}
If Assumptions \ref{as:1} - \ref{as:5} hold and $\omega_T$ satisfies the conditions $\lim_{T \to \infty} T \gamma_T / \omega_T = 0$ and $\lim_{T \to \infty} \omega_T/\underset{1 \leq i \leq m_0}{\min} \vert t^0_i - t^0_{i-1} \vert = 0$, then, as $T \to \infty$, $(\widehat{\widehat{m}}, \widehat{\widehat{\boldsymbol{t}}})$ satisfies
\begin{equation*}
P \left( \widehat{\widehat{m}} = m_0 \right) \to 1,
\end{equation*}
and it exists a constant $B > 0$ such that
\begin{equation*}
P \left( \underset{1 \leq i \leq m_0}{\max} \vert \widehat{\widehat{t}}_i - t^0_i \vert \leq B T \gamma_T \right) \to 1.
\end{equation*}
\end{theorem}

\begin{remark}
The conditions for $\omega_T$ given in the theorem, combined with the assumption that $\gamma_T/\lambda_T \to \infty$ as $T \to \infty$,  determine the penalty term of the IC. Through $\gamma_T$, the penalty term is linked to the tuning parameter of the group LASSO estimator. For example, using the sequence of tuning parameters found in Theorem \ref{th:1} and being of order $O( \log T / T )^{1/4}$, the conditions are satisfied for $\omega_T = C T^{3/4} \log T$ for some $C > 0$. For practical purposes, $C$ can be set in analogy to the construction of the BIC so that the second term in Equation~\eqref{eq:IC} penalizes the total number of nonzero coefficients.
\end{remark}

\begin{remark}
Theorem \ref{th:3} shows that the second step refinement leads to consistent estimation of the true number of breaks. The second property is stronger that the one expressed in Theorem \ref{th:2} and does not refer to the Hausdorff distance but shows that every breakpoint is identified within a $T \gamma_T$ neighborhood.
\end{remark}

If $|\mathcal{A}_T|$ is relatively large, evaluating every combination of breakpoints again becomes computationally intensive. Hence, we follow \cite{ChanYauZhang2014} and use a backward elimination algorithm to successively remove the most redundant breakpoint that corresponds to the largest reduction of the IC until no further improvement is possible. This means that including this breakpoint improves the fit sufficiently to outweigh the costs of estimating the coefficients for an additional regime. The details of the algorithm are outlined in Supplementary Material A.\footnote{Supplementary Material A can be found here: \url{https://karstenschweikert.github.io/mequ_ci/mequ_ci_suppA.pdf}}
We denote the set of estimated breakpoints obtained from the BEA by $\mathcal{A}^*_T = (\widehat{t}^*_1, \dots, \widehat{t}^*_{|\mathcal{A}^*_T|})$. The next theorem shows that the estimator based on the BEA has identical asymptotic properties. 

\begin{theorem}\label{th:4}
For the same conditions as in Theorem \ref{th:3}, it holds for $T \to \infty$ that
\begin{equation*}
P \left( |\mathcal{A}^*_T| = m_0 \right) \to 1,
\end{equation*}
and it exists a constant $B > 0$ such that
\begin{equation*}
P \left( \underset{1 \leq i \leq m_0}{\max} \vert \widehat{t}^*_i - t^0_i \vert \leq B T \gamma_T \right) \to 1.
\end{equation*}
\end{theorem}

Using the BEA, it is also possible to optimize another information criterion, say the BIC, for each regime to eliminate some variables from all equations. This allows us to investigate whether some variables lose importance during parts of the sample period. Depending on the chosen model structure this could even be interpreted as some variables dropping out of the long-run equilibrium relationship for a certain period.

Applying scaling factors to the integrated regressors and the linear trend in our model ensures that all regressors have the same order. In turn, this means that the OLS estimator in the second step, $\widehat{\widehat{\boldsymbol{\theta}}}_j$, has the same convergence rates for all coefficients in the model. In principle, it is also possible to conduct post-LASSO OLS estimation (after the second step) without scaling factors to benefit from higher convergence rates of the estimator for coefficients of trending variables.

\section{Simulation}\label{sec:sim}
\noindent
We conduct simulation experiments to assess the adequacy of our technical results presented in \autoref{sec:method}. Specifically, we investigate the finite sample performance of our estimator with respect to the speed and accuracy in finding the exact number of breaks and their location. In principle, we can optimize \eqref{eq:gfl2} directly using a coordinate descent algorithm like the one proposed in \cite{Breheny2015} for a grid of $\lambda_T$ values. However, such an algorithm is not computationally efficient for change-point problems. Thus, for the simulations and the empirical application, we rely on a modified group LARS algorithm that approximates the solution for the first step and the BEA for the second step. The choice of the tuning parameter $\lambda_T$ is translated to pre-specifying the maximum number of breakpoint candidates $M$, i.e.\ the maximum number of non-zero groups in $\hat{\boldsymbol{\theta}}(T)$ that is returned when the LARS implementation is used. The modified group LARS algorithm evaluates each point in time as a breakpoint candidate but returns only the $M$ most relevant breaks. According to Theorem 2, we know that the group LASSO estimator overselects breaks in the first step. Hence, $M$ should be set large enough to encompass all true breakpoints and some additional falsely selected non-zero groups. The BEA then asymptotically guarantees that the set of change-points is attained in the second step. Further, the minimum distance between breaks needs to be specified based on the number of coefficients in the model to guarantee that these coefficients can be estimated accurately in each regime. Details about the modified group LARS algorithm, the BEA, and additional simulation results are included in Supplementary Material A. 

We consider model specifications with one, two and four breakpoints, respectively. The following DGP is employed to model a multiple equations cointegrating regression with multiple structural breaks, 
\begin{equation}
\label{eq:mc.dgp}
\begin{array}{lllllll}
Y_t & = & A_t X_t + \delta_t t + \mu + B_t w_t + u_t, & & u_t & \sim & N(0,\Sigma_u), \\
X_t & = &  X_{t-1} + \xi_t, & & \xi_t & \sim & N(0,\Sigma_{\xi}), \\
w_t & = & \Phi w_{t-1} + e_t, & & e_t & \sim & N(0,\Sigma_e), \\
\end{array}
\end{equation}
where $X_t = (X_{1t}, X_{2t}, \dots, X_{Nt})'$, $\Sigma_{u,ii} = \sigma^2_u$, $\Sigma_{\xi,ii} = \sigma^2_{\xi}$ and $\Sigma_{e,ii} = \sigma^2_e$, for $i = 1, \dots, q$, i.e.\ the innovations of our generated processes have multivariate normal distributions with identical variances. We choose the variances of these processes so that each column of $Z_t = (T^{-1/2} X_t', T^{-1}t, w_t')'$ has variances less than or equal to the error term variances. To achieve this, we set the variances to $\sigma^2_{\xi} = \sigma^2_e = \sigma^2_u = 1$.
$\mu$ is a non-zero intercept vector, $A_t$ and $B_t$ are time-varying coefficient matrices with at least one non-zero entry in the baseline specification and a finite number of breaks. $\delta_t$ is a time-varying $q$-dimensional vector and $\Phi$ is a coefficient matrix for the VAR(1) process that fulfills the required stationarity conditions. For simplicity, we choose a diagonal matrix with each diagonal entry equal to 0.5. For the main results, we set $q = 2$ and use the following coefficient matrices:
\begin{equation}
A_{t_0} = B_{t_0} = \begin{bmatrix}
2 & 0\\
0 & 2
\end{bmatrix}, \qquad
A_{t_j} = A_{t_j-1} + c \begin{bmatrix}
2 & 0\\
0 & 2
\end{bmatrix}, \qquad
B_{t_j} = B_{t_j-1} + c \begin{bmatrix}
2 & 0\\
0 & 2
\end{bmatrix}, \quad
j = 1, \dots, m_0,
\label{eq:dgp_coefs}
\end{equation}
with $c = 1$ and the subscript $t_0$ denoting the initial coefficient matrix before the first breakpoint. Moreover, we set $\mu = (2,2)'$, $\delta_{t_0} = (2,2)'$ and $\delta_{t_j} = \delta_{t_j-1} + c (2,2)'$ for $j = 1, \dots, m_0$. In our baseline specification, the coefficient changes amount to two standard deviations of the error terms. Similar to \cite{BaiLumsdaineStock1998}, we also specify $c \in \lbrace 0.25, 0.5, 1.5 \rbrace$ to investigate the performance for smaller and larger break magnitudes. Moreover, we investigate the effect of cross-correlated errors, serially correlated errors, and endogenous regressors. The results of those robustness checks are included in Supplementary Material A to conserve space.

Naturally, the ability of all structural break estimators to detect breaks depends on the overall signal strength. \cite{NiuHaoZhang2015} define signal strength in change-point models by $S_{NHZ} = m_{\theta}^2 I_{\min}$, where $I_{\min} = \min_{1 \leq j \leq m_0+1} \vert t_j - t_{j-1} \vert$ is the minimum distance between breaks and $m_{\theta}$ is the minimum jump size as defined in Assumption \ref{as:5}. For our main simulations concerned with showing the consistency of the two-step estimator, we follow \cite{Schweikert2021} and use equal jump sizes for multiple breaks as well as locating the breaks with equidistant spacing between them. Hence, overall signal strength is a linear function of the sample size in our simulations. We choose a minimum of 50 observation per regime and double the sample sizes in line with the conventional asymptotics specified in \autoref{sec:first}. Consequently, the sample sizes chosen differ for an increasing number of breaks. Note that 12 coefficients are present in the full model specification which requires a substantial number of observations in each regime to estimate them precisely. 

In \autoref{tab:sb_sim_bea}, we report our results for $r = 2$ integrated regressors, $s = 2$ stationary regressors, a time trend, and $q = 2$ equations. We specify our model for one break located at $\tau = 0.5$, two breaks at $\tau = (0.33, 0.67)$ and four breaks at $\tau = (0.2, 0.4, 0.6, 0.8)$ to have an equidistant spacing on the unit interval. 
Since the choice of a change-point detection algorithm is often reflective of a trade-off between speed and accuracy, we first compute the average computing time in seconds (Time [s]) for each sample size.\footnote{All simulations experiments are conducted on a computer with an Intel i5-6500 CPU at 3.20GHz and 16GB RAM.} Next, we compute the percentages of correct estimation (pce) of the true number of breaks $m_0$ and measure the accuracy of the break date estimation conditional on the correct estimation of $m_0$. For this matter, we compute the standard deviations of the estimated relative timing. The estimated coefficients are not reported to conserve space but can be obtained from the author upon request.

Our simulation results reveal that the two-step estimator (panel A) is less precise compared with the likelihood-based approach (panel B) when it comes to the estimation of the individual break locations. This is not surprising considering that the likelihood-based approach according to \cite{QuPerron2007} rests on a dynamic programming algorithm and uses repeated OLS regressions to determine the location of the breaks in an almost exact fashion. The trade-off here is then computing time versus precision. As outlined above, the computational efficiency of the two-step estimator is much higher ($O(M^3 +MT)$ versus $O(T^2)$) which manifests in reduced computing times at moderate to large samples. For example, while it takes the likelihood-approach several minutes (on average 2,013 seconds) on a modern computer to solve the change-point problem for $T=2,000$, the two-step estimator solves the same problem in seconds (on average 36 seconds). It is also important to note that, in contrast to the two-step estimator, the likelihood-based approach is conceptually not able to consistently estimate the exact number of breaks. As expected, the sequential test procedure reaches its specified nominal confidence level at large samples sizes and in our setting ($\alpha = 0.05$) detects the number of breaks in roughly 95\% of all cases.\footnote{It seems to be undersized for small sample sizes and a larger number of breaks, reaching a 100\% detection rate for $m = 4$ and $T = 250$ but then declining to 95\% for larger samples.} Instead, the two-step estimator already attains close to a 100\% detection rate at moderate sample sizes. This also means that the most difficult cases (leading to a non-rejection of the sequential test's hypothesis) are not considered for the evaluation of the precision of the likelihood-based approach because in columns two to six of \autoref{tab:sb_sim_bea} only those cases with the correctly estimated number of breaks can be properly evaluated. \cite{BaiPerron1998} suggest to let the size of those sequential tests go to zero asymptotically to avoid misspecification. However, to implement this in practice is difficult without any guidance on the exact rate for an adjusted $\alpha$. Alternatively, an information criterion might be used to compare specifications with a different number of breaks.

We further study the performance of the two-step estimator in settings with structural breaks of smaller or larger magnitude.\footnote{The results can be found in Tables S1 -- S3 in Supplementary Material A.} For instance, we apply the factor $c=1.5$ to increase the break magnitude in Equation~\eqref{eq:dgp_coefs} to three standard deviations of the error term. Although we find a slightly better detection of the number of breaks, we observe no substantial effect on the precision in terms of finding the true location of the breaks. It appears that increasing the magnitude does not further improve the performance in small samples. A reason for this upper bound in precision is the pre-selection of breakpoint candidates in the first step which is only accurate up to a $T\gamma_T$ neighborhood of the true breakpoints. In contrast, reducing the magnitude when applying the factor $c=0.5$, i.e., to one standard deviation, leads to worse results for small sample sizes. For example, in the case of four active break and $T=250$ observations, we find that the rate of correctly detecting the number of breaks drops from 89.0\% in the baseline specification to 68.3\%. However, when the sample size increases to $T=500$, we almost reach the same rate of detection and a similar precision for the break's location. The likelihood-based approach is still very precise when the break magnitudes are reduced by the factor $c=0.5$. However, when we reduce them further by setting $c=0.25$, the detection rate breaks down to, e.g., 6.7\% for four breaks and $T=250$ observations while the two-step approach still reaches a 40.1\% detection rate. We conclude that the performance of the two-step approach naturally depends on the size of the break magnitude but it is fairly robust in medium to large sample sizes which should be its primary field of application considering its improvements in terms of computational costs.

To investigate whether the results are driven by the integrated regressors or the linear trend in the model, we run the simulation experiments for a reduced SUR model specification. Here, we generate data under the restriction that $A_i = diag(0)$ and $\delta_i = 0$ for $i = 0, \dots m_0$. The results are reported in \autoref{tab:sb_sim_sur}. We consider two variants of the simulation experiment: first with no cross-correlation between the error terms for both equations as in Equation~\eqref{eq:mc.dgp} and second with cross-correlation coefficient $\rho = 0.5$. We find almost identical precision for the SUR model and no substantial difference in the presence of moderate cross-correlation. This shows that the precision of the two-step approach is predominately driven by the magnitude of breaks in terms of their Euclidean distance for the vector of coefficients and the magnitude is the same for the SUR specification. The reduced number of coefficients in the SUR model does not seem to improve the performance substantially, because the sample size in each regime is already large enough to estimate the full model. Additional simulation experiments (not reported) show that this aspect certainly gains importance for smaller regimes with less observations.

We extend the model with a third equation to investigate if the results either improve because a common break is indicated in another regression equation \citep[similar to results obtained in][]{BaiLumsdaineStock1998, QuPerron2007} or deteriorate because the detection relies on additional coefficient changes that need to be estimated. We consider a special case in which the break magnitudes stay constant after adding the third equation. In practice, we hope that estimating the structural breaks jointly in all equations includes some larger coefficient changes that help to find the common break dates. The results for the $q=3$ case, reported in \autoref{tab:sb_sim_q3}, show that we reach similar detection rates and approximately the same precision for the timing of the breaks. As the break magnitudes are identical for each coefficient, the $q=3$ setting is a straightforward extension of the $q=2$ setting. To investigate further whether breaks in a subset of the coefficients can be reliably detected in a larger system, we consider a partial break specification in the $q=2$ case. Here, only the coefficients of the first equation change. The results are reported in Table S5 in Supplementary Material A and again show that the two-step estimator's performance is mostly determined by the total break magnitude which is slightly reduced in this case.

Finally, we turn to the one-break case of our main specification again and move the break to the right boundary of the unit interval. We fix the number of observations in the second regime to 25 and increase the overall sample size. Consequently the breakpoint moves further to the boundary each time we double the sample size. The true break fractions for the sample sizes $T \in \lbrace 100, 200, 400, 800 \rbrace$ are 0.75, 0.875, 0.9375, and 0.96875, respectively. We also consider a two-break setting where the middle regime shrinks relative to the total sample size. We set the first breakpoint at $\tau_1 = 0.5$ and the second one at $\tau_2 = 0.5 + 25/T$. The results for both experiments are reported in \autoref{tab:sb_sim_regimes}. These exercises show that the two-step estimator is robust to setting a small minimum break distance while the sequential tests to detect breaks in the likelihood-based approach require a rather large trimming parameter to ensure the right size and sufficient power. Of course, if the trimming parameter is set to 0.15 as it is suggested for empirical data \citep{QuPerron2007}, it is infeasible to detect breaks that are too close to each other or near the boundary of the unit interval. In case of one break at the boundary, the likelihood-based estimator in our simulations (almost) always falsely indicates a break at 0.85.

\begin{table}[htbp]
\caption{Estimation of (multiple) structural breaks in the full model}
\begin{center}
\scalebox{0.75}{
\begin{tabular}{c c c c c c c c c c c c c c c c c c c c c c c c c c c c c c}
\toprule[1pt]
 &  \multicolumn{14}{l}{Panel A: Group LASSO with BEA} \\ 
\cmidrule{2-7}
 & & & & & & & & & & & & \\ 
 &  \multicolumn{14}{l}{SB1: ($\tau = 0.5$)} \\ 
$T$ & & Time [s] & & $pce$ &  & $\tau$ &  &  &  &  &  &  &  & \\ 
\midrule[0.5pt]
100 & & 0.04 & & 98.9 & & 0.501 (0.014) & &  & &  & &  & & \\
200 & & 0.08 & & 100 & & 0.500 (0.007) & &  & &  & &  & & \\
400 & & 0.26 & & 100 & & 0.500 (0.003) & &  & &  & &  & & \\
800 & & 1.64 & & 100 & & 0.500 (0.001) & &  & &  & &  & & \\
 & & & & & & & & & & & & \\ 
 &  \multicolumn{14}{l}{SB2: ($\tau_1 = 0.33$, $\tau_2 = 0.67$)} \\ 
$T$ & & Time [s] & & $pce$ &  & $\tau_1$ &  & $\tau_2$ &  &  &  &  &  &  &  &  &  &  &  &  \\
\midrule[0.5pt]
150 & & 0.08 & & 97.6 &  & 0.335 (0.030) &  & 0.659 (0.026) &  &  &  &  &  &  &  &  &  &  &  &  \\
300 & & 0.25 & & 100 &  & 0.333 (0.018) &  & 0.666 (0.014) &  &  &  &  &  &  &  &  &  &  &  &  \\
600 & & 1.53 & & 100 &  & 0.331 (0.009) &  & 0.668 (0.007) &  &  &  &  &  &  &  &  &  &  &  &  \\
1200 & & 10.65 & & 100 &  & 0.331 (0.004) &  & 0.669 (0.003) &  &  &  &  &  &  &  &  &  &  &  &  \\
 & & & & & & & & & & & & \\ 
 &  \multicolumn{14}{l}{SB4: ($\tau_1 = 0.2$, $\tau_2 = 0.4$, $\tau_3 = 0.6$, $\tau_4 = 0.8$)} \\ 
$T$ & & Time [s] & & $pce$ &  & $\tau_1$ &  & $\tau_2$ &  & $\tau_3$ &  & $\tau_4$\\
\midrule[0.5pt]
250 & & 0.17 & & 89.0 & & 0.217 (0.030) & & 0.404 (0.022) & & 0.596 (0.019) & & 0.788 (0.028)\\
500 & & 0.74 & & 98.2 & & 0.203 (0.017) & & 0.402 (0.012) & & 0.598 (0.009) & & 0.803 (0.012)\\
1000 & & 4.68 & & 99.9 & & 0.199 (0.008) & & 0.401 (0.006) & & 0.599 (0.005) & & 0.800 (0.008)\\
2000 & & 36.25 & & 100 & & 0.200 (0.003) & & 0.401 (0.003) & & 0.599 (0.002) & & 0.800 (0.003)\\
 & & & & & & & & & & & & \\ 
 &  \multicolumn{14}{l}{Panel B: Likelihood-based approach} \\ 
\cmidrule{2-7}
 & & & & & & & & & & & & \\ 
 &  \multicolumn{14}{l}{SB1: ($\tau = 0.5$)} \\ 
$T$ & & Time [s] & & $pce$ &  & $\tau$ &  &  &  &  &  &  &  & \\ 
\midrule[0.5pt]
100 & & 1.22 & & 91.3 & & 0.499 (0.041) & &  & &  & &  & & \\
200 & & 4.99 & & 93.0 & & 0.500 (0.010) & &  & &  & &  & & \\
400 & & 18.71 & & 94.5 & & 0.500 (0.005) & &  & &  & &  & & \\
800 & & 89.07 & & 94.7 & & 0.500 (0.003) & &  & &  & &  & & \\
 & & & & & & & & & & & & \\ 
 &  \multicolumn{14}{l}{SB2: ($\tau_1 = 0.33$, $\tau_2 = 0.67$)} \\ 
$T$ & & Time [s] & & $pce$ &  & $\tau_1$ &  & $\tau_2$ &  &  &  &  &  &  &  &  &  &  &  &  \\
\midrule[0.5pt]
150 & & 4.61 & & 94.0 &  & 0.327 (0.005) &  & 0.667 (0.004) &  &  &  &  &  &  &  &  &  &  &  &  \\
300 & & 15.28 & & 95.0 &  & 0.330 (0.002) &  & 0.670 (0.002) &  &  &  &  &  &  &  &  &  &  &  &  \\
600 & & 60.16 & & 96.1 &  & 0.330 (0.001) &  & 0.670 (0.001) &  &  &  &  &  &  &  &  &  &  &  &  \\
1200 & & 317.78 & & 95.5 &  & 0.330 (0.001) &  & 0.670 (0.001) &  &  &  &  &  &  &  &  &  &  &  &  \\
 & & & & & & & & & & & & \\ 
 &  \multicolumn{14}{l}{SB4: ($\tau_1 = 0.2$, $\tau_2 = 0.4$, $\tau_3 = 0.6$, $\tau_4 = 0.8$)} \\ 
$T$ & & Time [s] & & $pce$ &  & $\tau_1$ &  & $\tau_2$ &  & $\tau_3$ &  & $\tau_4$\\
\midrule[0.5pt]
250 & & 16.59 & & 99.8 & & 0.200 (0.004) & & 0.400 (0.004) & & 0.600 (0.004) & & 0.800 (0.004)\\
500 & & 64.93 & & 96.7 & & 0.200 (0.002) & & 0.400 (0.001) & & 0.600 (0.001) & & 0.800 (0.001)\\
1000 & & 322.58 & & 95.6 & & 0.200 (0.001) & & 0.400 (0.001) & & 0.600 (0.001) & & 0.800 (0.001)\\
2000 & & 2013.69 & & 95.1 & & 0.200 (0.001) & & 0.400 (0.000) & & 0.600 (0.000) & & 0.800 (0.000)\\
\bottomrule[1pt]
\end{tabular}
}
\end{center}
\label{tab:sb_sim_bea}
\begin{tablenotes}
\scriptsize
\item Note: We use 1,000 replications of the data-generating process given in Equation~\eqref{eq:mc.dgp}. The variance of the error terms is $\sigma^2_{\xi} = \sigma^2_e = \sigma^2_u = 1$. The first subpanel reports the results for one active breakpoint at $\tau = 0.5$, the second subpanel considers two active breakpoints at $\tau_1 = 0.33$ and $\tau_2 = 0.67$ and the third subpanel has four active breakpoints at $\tau_1 = 0.2$, $\tau_2 = 0.4$, $\tau_3 = 0.6$, and $\tau_4 = 0.8$. Time [s] denotes the average computing time for one replication in seconds. Standard deviations are given in parentheses. We conduct the $\sup(l+1|l)$ test at the 5\% level to determine the number of breaks. Critical values are provided in \cite{QuPerron2007}.
\end{tablenotes}
\end{table}

\begin{table}[htbp]
\caption{Estimation of (multiple) structural breaks in the SUR model specification using the group LASSO with BEA}
\begin{center}
\scalebox{0.75}{
\begin{tabular}{c c c c c c c c c c c c c c c c c c c c c c c c c c c c}
\toprule[1pt]
 &  \multicolumn{14}{l}{Panel A: No cross-correlation} \\ 
\cmidrule{2-7}
 & & & & & & & & & & & & \\ 
 &  \multicolumn{14}{l}{SB1: ($\tau = 0.5$)} \\ 
$T$ & & Time [s] & & $pce$ &  & $\tau$ &  &  &  &  &  &  &  & \\ 
\midrule[0.5pt]
100 & & 0.03 & & 100 & & 0.500 (0.010) & &  & &  & &  & & \\
200 & & 0.06 & & 100 & & 0.500 (0.005) & &  & &  & &  & & \\
400 & & 0.20 & & 100 & & 0.500 (0.002) & &  & &  & &  & & \\
800 & & 1.27 & & 100 & & 0.500 (0.001) & &  & &  & &  & & \\
 & & & & & & & & & & & & \\ 
 &  \multicolumn{14}{l}{SB2: ($\tau_1 = 0.33$, $\tau_2 = 0.67$)} \\ 
$T$ & & Time [s] & & $pce$ &  & $\tau_1$ &  & $\tau_2$ &  &  &  &  &  &  &  &  &  &  &  &  \\
\midrule[0.5pt]
150 & & 0.04 & & 98.9 &  & 0.337 (0.034) &  & 0.656 (0.033) &  &  &  &  &  &  &  &  &  &  &  &  \\
300 & & 0.23 & & 100 &  & 0.333 (0.019) &  & 0.667 (0.019) &  &  &  &  &  &  &  &  &  &  &  &  \\
600 & & 1.39 & & 100 &  & 0.333 (0.009) &  & 0.668 (0.009) &  &  &  &  &  &  &  &  &  &  &  &  \\
1200 & & 9.70 & & 100 &  & 0.331 (0.005) &  & 0.669 (0.005) &  &  &  &  &  &  &  &  &  &  &  &  \\
 & & & & & & & & & & & & \\ 
 &  \multicolumn{14}{l}{SB4: ($\tau_1 = 0.2$, $\tau_2 = 0.4$, $\tau_3 = 0.6$, $\tau_4 = 0.8$)} \\ 
$T$ & & Time [s] & & $pce$ &  & $\tau_1$ &  & $\tau_2$ &  & $\tau_3$ &  & $\tau_4$\\
\midrule[0.5pt]
250 & & 0.15 & & 87.2 & & 0.215 (0.031) & & 0.406 (0.021) & & 0.594 (0.020) & & 0.786 (0.031)\\
500 & & 0.63 & & 99.9 & & 0.201 (0.013) & & 0.402 (0.012) & & 0.597 (0.011) & & 0.800 (0.013)\\
1000 & & 10.01 & & 100 & & 0.199 (0.008) & & 0.401 (0.006) & & 0.599 (0.005) & & 0.801 (0.008)\\
2000 & & 72.30 & & 100 & & 0.200 (0.003) & & 0.400 (0.003) & & 0.599 (0.003) & & 0.800 (0.003)\\
 & & & & & & & & & & & & \\ 
 &  \multicolumn{14}{l}{Panel B: Cross-correlated errors ($\rho = 0.5$)} \\ 
\cmidrule{2-7}
 & & & & & & & & & & & & \\ 
  &  \multicolumn{14}{l}{SB1: ($\tau = 0.5$)} \\ 
$T$ & & Time [s] & & $pce$ &  & $\tau$ &  &  &  &  &  &  &  & \\ 
\midrule[0.5pt]
100 & & 0.03 & & 100 & & 0.500 (0.010) & &  & &  & &  & & \\
200 & & 0.06 & & 100 & & 0.500 (0.005) & &  & &  & &  & & \\
400 & & 0.22 & & 100 & & 0.500 (0.002) & &  & &  & &  & & \\
800 & & 1.36 & & 100 & & 0.500 (0.001) & &  & &  & &  & & \\
 & & & & & & & & & & & & \\ 
 &  \multicolumn{14}{l}{SB2: ($\tau_1 = 0.33$, $\tau_2 = 0.67$)} \\ 
$T$ & & Time [s] & & $pce$ &  & $\tau_1$ &  & $\tau_2$ &  &  &  &  &  &  &  &  &  &  &  &  \\
\midrule[0.5pt]
150 & & 0.05 & & 97.8 &  & 0.337 (0.034) &  & 0.656 (0.033) &  &  &  &  &  &  &  &  &  &  &  &  \\
300 & & 0.22 & & 99.8 &  & 0.334 (0.019) &  & 0.667 (0.018) &  &  &  &  &  &  &  &  &  &  &  &  \\
600 & & 1.68 & & 100 &  & 0.332 (0.009) &  & 0.668 (0.009) &  &  &  &  &  &  &  &  &  &  &  &  \\
1200 & & 9.99 & & 100 &  & 0.332 (0.005) &  & 0.669 (0.005) &  &  &  &  &  &  &  &  &  &  &  &  \\
 & & & & & & & & & & & & \\ 
 &  \multicolumn{14}{l}{SB4: ($\tau_1 = 0.2$, $\tau_2 = 0.4$, $\tau_3 = 0.6$, $\tau_4 = 0.8$)} \\ 
$T$ & & Time [s] & & $pce$ &  & $\tau_1$ &  & $\tau_2$ &  & $\tau_3$ &  & $\tau_4$\\
\midrule[0.5pt]
250 & & 0.17 & & 93.1 & & 0.216 (0.031) & & 0.406 (0.020) & & 0.594 (0.020) & & 0.785 (0.031)\\
500 & & 0.58 & & 99.8 & & 0.200 (0.013) & & 0.402 (0.011) & & 0.597 (0.011) & & 0.800 (0.013)\\
1000 & & 12.26 & & 100 & & 0.199 (0.008) & & 0.401 (0.006) & & 0.599 (0.005) & & 0.801 (0.008)\\
2000 & & 78.36 & & 100 & & 0.200 (0.003) & & 0.400 (0.003) & & 0.599 (0.003) & & 0.800 (0.003)\\
\bottomrule[1pt]
\end{tabular}
}
\end{center}
\label{tab:sb_sim_sur}
\begin{tablenotes}
\scriptsize
\item Note: We use 1,000 replications of the data-generating process given in Equation~\eqref{eq:mc.dgp}. The variance of the error terms is $\sigma^2_{\xi} = \sigma^2_e = \sigma^2_u = 1$. The first subpanel reports the results for one active breakpoint at $\tau = 0.5$, the second subpanel considers two active breakpoints at $\tau_1 = 0.33$ and $\tau_2 = 0.67$ and the third subpanel has four active breakpoints at $\tau_1 = 0.2$, $\tau_2 = 0.4$, $\tau_3 = 0.6$, and $\tau_4 = 0.8$. Time [s] denotes the average computing time for one replication in seconds. Standard deviations are given in parentheses.
\end{tablenotes}
\end{table}

\begin{table}[htbp]
\caption{Estimation of (multiple) structural breaks in the full model ($q=3$) using the group LASSO with BEA}
\begin{center}
\scalebox{0.75}{
\begin{tabular}{c c c c c c c c c c c c c c c c c c c c c c c c c c c c}
\toprule[1pt]
 &  \multicolumn{14}{l}{SB1: ($\tau = 0.5$)} \\ 
$T$ & & Time [s] & & $pce$ &  & $\tau$ &  &  &  &  &  &  &  & \\ 
\midrule[0.5pt]
100 & & 0.04 & & 99.8 & & 0.500 (0.014) & &  & &  & &  & & \\
200 & & 0.08 & & 100 & & 0.500 (0.005) & &  & &  & &  & & \\
400 & & 0.35 & & 100 & & 0.500 (0.002) & &  & &  & &  & & \\
800 & & 2.15 & & 100 & & 0.500 (0.001) & &  & &  & &  & & \\
 & & & & & & & & & & & & \\ 
 &  \multicolumn{16}{l}{SB2: ($\tau_1 = 0.33$, $\tau_2 = 0.67$)} \\ 
$T$ & & Time [s] & & $pce$ &  & $\tau_1$ &  & $\tau_2$ &  &  &  &  &  &  &  &  &  &  &  &  \\
\midrule[0.5pt]
150 & & 0.08 & & 98.1 &  & 0.333 (0.032) &  & 0.663 (0.025) &  &  &  &  &  &  &  &  &  &  &  &  \\
300 & & 0.28 & & 100 &  & 0.333 (0.017) &  & 0.667 (0.014) &  &  &  &  &  &  &  &  &  &  &  &  \\
600 & & 1.72 & & 100 &  & 0.333 (0.008) &  & 0.668 (0.007) &  &  &  &  &  &  &  &  &  &  &  &  \\
1200 & & 14.49 & & 100 &  & 0.331 (0.004) &  & 0.669 (0.004) &  &  &  &  &  &  &  &  &  &  &  &  \\
 & & & & & & & & & & & & \\ 
 &  \multicolumn{14}{l}{SB4: ($\tau_1 = 0.2$, $\tau_2 = 0.4$, $\tau_3 = 0.6$, $\tau_4 = 0.8$)} \\ 
$T$ & & Time [s] & & $pce$ &  & $\tau_1$ &  & $\tau_2$ &  & $\tau_3$ &  & $\tau_4$\\
\midrule[0.5pt]
250 & & 0.18 & & 89.3 & & 0.217 (0.031) & & 0.405 (0.021) & & 0.597 (0.019) & & 0.789 (0.029)\\
500 & & 0.85 & & 97.7 & & 0.202 (0.016) & & 0.402 (0.011) & & 0.597 (0.009) & & 0.802 (0.013)\\
1000 & & 12.25 & & 99.2 & & 0.199 (0.008) & & 0.401 (0.006) & & 0.599 (0.005) & & 0.800 (0.008)\\
2000 & & 103.64 & & 99.8 & & 0.200 (0.003) & & 0.401 (0.003) & & 0.599 (0.002) & & 0.800 (0.003)\\
\bottomrule[1pt]
\end{tabular}
}
\end{center}
\label{tab:sb_sim_q3}
\begin{tablenotes}
\scriptsize
\item Note: We use 1,000 replications of a variant of the data-generating process given in Equation~\eqref{eq:mc.dgp} with an additional equation but the same break magnitudes. The variance of the error terms is $\sigma^2_{\xi} = \sigma^2_e = \sigma^2_u = 1$. The first subpanel reports the results for one active breakpoint at $\tau = 0.5$, the second subpanel considers two active breakpoints at $\tau_1 = 0.33$ and $\tau_2 = 0.67$ and the third subpanel has four active breakpoints at $\tau_1 = 0.2$, $\tau_2 = 0.4$, $\tau_3 = 0.6$, and $\tau_4 = 0.8$. Time [s] denotes the average computing time for one replication in seconds. Standard deviations are given in parentheses.
\end{tablenotes}
\end{table}

\begin{table}[htbp]
\caption{Estimation of (multiple) structural breaks in the full model with shrinking regimes}
\begin{center}
\scalebox{0.75}{
\begin{tabular}{c c c c c c c c c c c c c c c c c c c c c c c c c c c c c c}
\toprule[1pt]
 &  \multicolumn{14}{l}{Panel A: Group LASSO with BEA} \\ 
\cmidrule{2-7}
 & & & & & & & & & & & & \\ 
 &  \multicolumn{14}{l}{SB1: ($\tau = 1 - 25/T$)} \\ 
$T$ & & Time [s] & & $pce$ &  & $\tau$ &  &  &  &  &  &  &  & \\ 
\midrule[0.5pt]
100 & & 0.03 & & 100 & & 0.747 (0.019) & &  & &  & &  & & \\
200 & & 0.07 & & 99.5 & & 0.873 (0.010) & &  & &  & &  & & \\
400 & & 0.29 & & 99.7 & & 0.936 (0.006) & &  & &  & &  & & \\
800 & & 1.69 & & 99.8 & & 0.967 (0.004) & &  & &  & &  & & \\
 & & & & & & & & & & & & & & \\ 
 &  \multicolumn{16}{l}{SB2: ($\tau_1 = 0.5$, $\tau_2 = 0.5 + 25/T$)} \\ 
$T$ & & Time [s] & & $pce$ &  & $\tau_1$ &  & $\tau_2$ &  &  &  &  &  &  &  &  &  &  &  &  \\
\midrule[0.5pt]
150 & & 0.09 & & 84.2 &  & 0.505 (0.019) &  & 0.661 (0.022) &  &  &  &  &  &  &  &  &  &  &  &  \\
300 & & 0.26 & & 91.9 &  & 0.500 (0.017) &  & 0.584 (0.016) &  &  &  &  &  &  &  &  &  &  &  &  \\
600 & & 1.59 & & 94.1 &  & 0.499 (0.009) &  & 0.542 (0.008) &  &  &  &  &  &  &  &  &  &  &  &  \\
 & & & & & & & & & & & & & & \\ 
 &  \multicolumn{14}{l}{Panel B: Likelihood-based approach} \\ 
\cmidrule{2-7}
 & & & & & & & & & & & & \\ 
 &  \multicolumn{14}{l}{SB1: ($\tau = 1 - 25/T$)} \\ 
$T$ & & Time [s] & & $pce$ &  & $\tau$ &  &  &  &  &  &  &  & \\ 
\midrule[0.5pt]
100 & & 1.40 & & 96.7 & & 0.750 (0.006) & &  & &  & &  & & \\
200 & & 3.65 & & 97.3 & & 0.854 (0.004) & &  & &  & &  & & \\
400 & & 19.72 & & 98.5 & & 0.851 (0.005) & &  & &  & &  & & \\
800 & & 103.17 & & 98.7 & & 0.850 (0.003) & &  & &  & &  & & \\
 & & & & & & & & & & & & & & \\ 
 &  \multicolumn{16}{l}{SB2: ($\tau_1 = 0.5$, $\tau_2 = 0.5 + 25/T$)} \\ 
$T$ & & Time [s] & & $pce$ &  & $\tau_1$ &  & $\tau_2$ &  &  &  &  &  &  &  &  &  &  &  &  \\
\midrule[0.5pt]
150 & & 3.75 & & 94.3 &  & 0.505 (0.019) &  & 0.661 (0.022) &  &  &  &  &  &  &  &  &  &  &  &  \\
300 & & 14.88 & & 95.5 &  & 0.463 (0.034) &  & 0.615 (0.034) &  &  &  &  &  &  &  &  &  &  &  &  \\
600 & & 72.04 & & 96.3 &  & 0.443 (0.056) &  & 0.596 (0.057) &  &  &  &  &  &  &  &  &  &  &  &  \\
\bottomrule[1pt]
\end{tabular}
}
\end{center}
\label{tab:sb_sim_regimes}
\begin{tablenotes}
\scriptsize
\item Note: We use 1,000 replications of the data-generating process given in Equation (10). The variance of the error terms is $\sigma^2_{\xi} = \sigma^2_e = \sigma^2_u = 1$. The first subpanel reports the results for the two step estimator with one active breakpoint at $\tau = 1 - 25/T$, the second subpanel reports the results for two active breakpoints at $\tau_1 = 0.5$, $\tau_2 = 0.5 + 25/T$. The likelihood based approach is applied with trimming parameter $0.15$. Time [s] denotes the average computing time for one replication in seconds. Standard deviations are given in parentheses. We conduct the $\sup(l+1|l)$ test at the 5\% level to determine the number of breaks. Critical values are provided in \cite{QuPerron2007}.
\end{tablenotes}
\end{table}

\section{Empirical Application}\label{sec:emp}
\noindent
In our empirical application, we apply the two-step estimator to US term structure data. Thereby, we revisit the study by \cite{Hansen2003} who proposes a framework to test for structural change in cointegrated vector autoregressions. The author tests two potential structural breaks in September 1979 and October 1982 that coincide with large changes in the Fed's policy. Only after accounting for these structural changes, the long-run implication of the expectations hypothesis (EHT) cannot be rejected. In the subsequent analysis, we study more recent data on the US term structure and detect structural breaks without assuming any prior knowledge about their location.

Following \cite{Campbell1987}, we expect that the term structure of interest rates is determined by the expected future spot rates being equal to the future rate plus a time-invariant term premium in the long-run. This implies that, independent of the maturity, the yields should be cointegrated with pairwise cointegrating vector $(1,-1)$. However, several early studies report that the EHT fails in empirical practice \citep[see, for example,][]{Froot1989, Campbell1991}. Besides other empirical difficulties, structural breaks are named as one of the important reasons for this failure \citep{Lanne1999, Sarno2007, Bulkley2011}. Several studies investigate whether regime shifts in the term structure of interest rates are related to changes in monetary policy \citep{Tillmann2007, Thornton2018}. The important question for applied researchers and policy-makers is whether these equilibrium relationships are robust over different regimes.

\begin{figure}[ht!]
\centering
\includegraphics[scale=0.65]{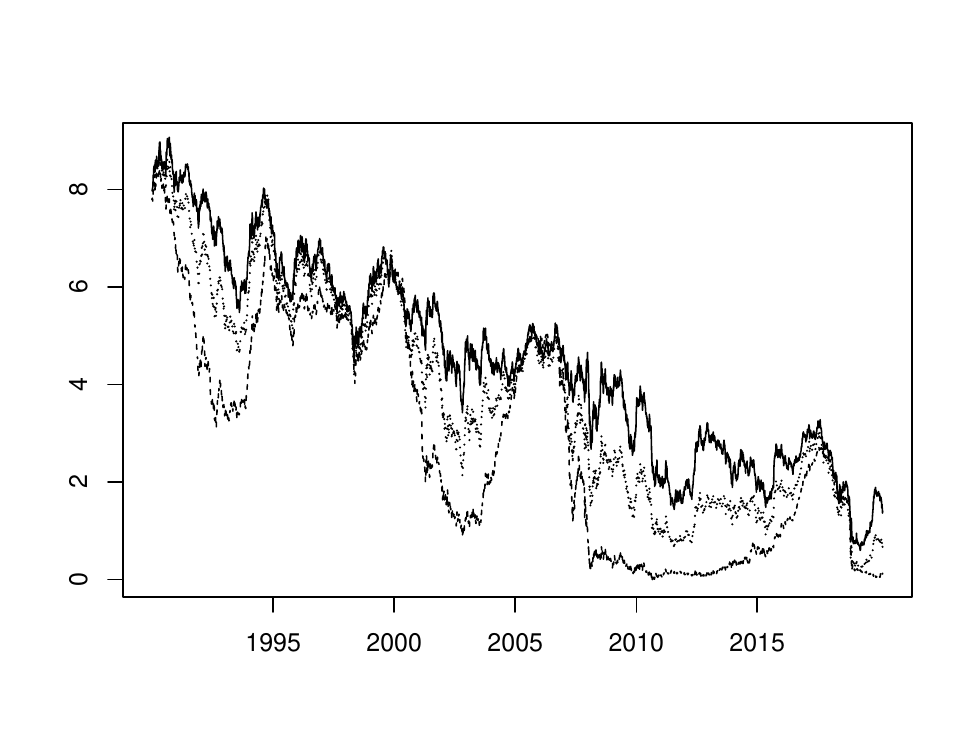}
\captionof{figure}{\small{Fitted yields (in \%) on 10-year (solid), 5-year (dotted, and 1-year (dashed) zero coupon US bonds.}}
\label{fig:tsplot}
\end{figure}

Using daily data from January 1990 to July 2021 on the term structure of US interest rates, we end up with more than 8,000 observations to estimate the term structure model. We again emphasize that almost exact segmentation algorithms like the one used for the likelihood-based approach are substantially slower than the two-step procedure based on the group LASSO estimator (26,380 seconds versus 25 seconds for solving the change point problem). Taking into account that several re-estimations of the model must be conducted to find the right specification and perform robustness checks, a reduced computational burden is important to encourage routine checks for structural breaks in multivariate systems. We use fitted yields on zero coupon US bonds with 10-year ($r_{10y,t}$), 5-year ($r_{5y,t})$, and 1-year ($r_{1y,t}$) maturity in the term structure model,
\begin{eqnarray}
\label{eq:term}
          r_{10y,t} &=& \mu_1 + \beta_1 r_{1y,t} + u_{1,t} \\
\notag    r_{5y,t} &=& \mu_2 + \beta_2 r_{1y,t} + u_{2,t}.  
\end{eqnarray}
We choose a model specification matching the long-run component in the cointegrated VAR used in \cite{Hansen2003}. Each equation models the pairwise relationship between the longer term maturity and the short term maturity (a one-to-one relationship under the EHT), while the constant accounts for a term premium. Additional maturities could be analyzed, leading to additional equations in the model, but we try to maintain a simple model structure. The data are produced according to the approach of \cite{Kim2005} fitting a simple three-factor arbitrage-free term structure model to U.S. Treasury yields since 1990, in order to evaluate the behavior of long-term yields, distant-horizon forward rates, and term premiums.\footnote{The data can be downloaded from the St Louis Fed's database FRED.} \autoref{fig:tsplot} provides a time series plot of the data. We assume that the individual variables follow unit root processes.\footnote{See the discussion in \cite{Hansen2003} and the references given therein why this assumption is useful for the empirical modelling of the term structure although some features like the non-negativity of interest rates are arguments against it. We also conduct unit root tests which do not provide evidence against this hypothesis in this specific sample period.} The results of a Johansen trace test for the full sample suggest that the trivariate system is cointegrated with cointegration rank one or two depending on the specification of the deterministic terms. According to the long-run implications of the EHT, we would expect a cointegration rank of two but it is well-known that the Johansen trace tests are not robust to structural breaks in the cointegrating vectors and the deterministic terms \citep[see, for example,][]{Lutkepohl2004, Saikkonen2006}.

\begin{figure}[ht!]
\centering
\includegraphics[scale=0.65]{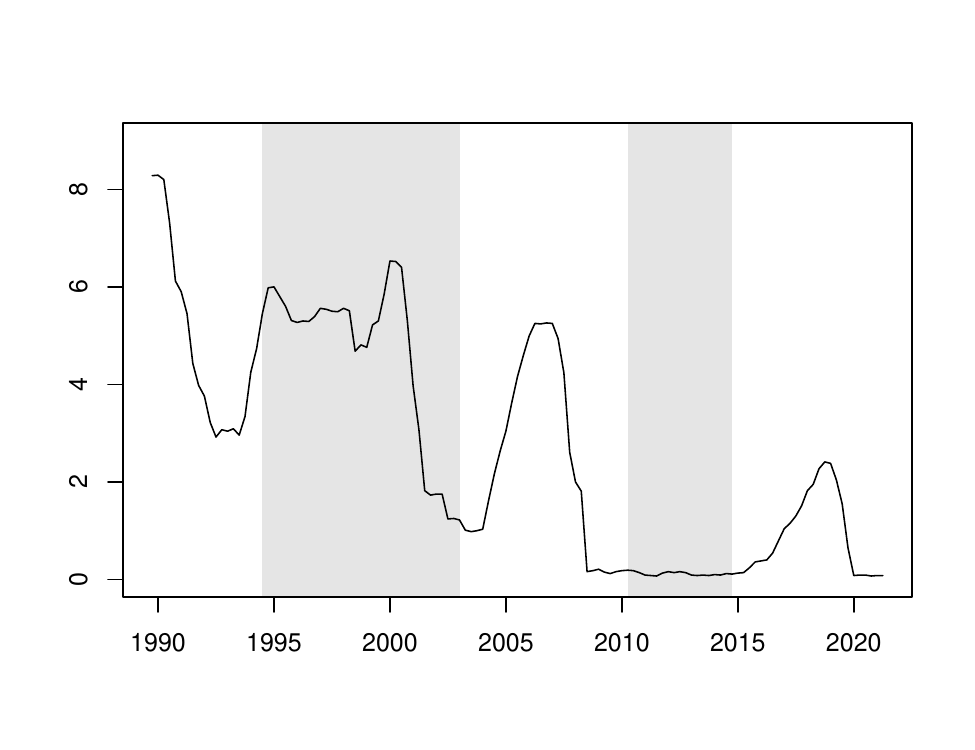}
\captionof{figure}{\small{Effective federal funds rate at a quarterly frequency. White and grey-shaded spaces mark the identified term structure regimes.}}
\label{fig:ffr.plot}
\end{figure}

We estimate the cointegrated term structure regression in Equation~\eqref{eq:term} with a dynamic OLS specification adding two leads and lags of $\Delta r_{1y,t}$. First, we estimate the model for the full sample without accounting for any structural breaks. The coefficient estimates are $\hat{\beta}_1 = 0.764 (0.116)$ and $\hat{\beta}_2 = 0.894 (0.078)$ which lead to a rejection of the EHT at the 5\% significance level for the 10-year maturity but not for the 5-year maturity.\footnote{Bootstrap standard errors are computed based on 600 replications of the sieve bootstrap method for cointegrating regressions proposed in \cite{Chang2006}.} In a second step, we try to capture all relevant structural breaks. Due to the large number of observations, we pre-specify a large maximum number of breaks, $M=40$, and maintain a minimum break distance of two month (50 daily observations) to obtain accurate coefficient estimates in each regime.\footnote{The results are robust for different choices of $M$ as long as $M > 4$. Since a larger value of $M$ allows the modified group LARS algorithm to take more steps, a larger value of $M$ can, in principle, positively affect the precision of the estimated break locations. However, the estimates do not change for $M > 40$.} We consider a specification with a constant and apply the required scaling factors so that each regressor has the same order.\footnote{The results with and without linear trend do not differ substantially.} Using the two-step group LASSO estimator, we obtain four structural breaks. All breakpoint estimates can be related to important monetary policy events. To show that, we depict the trajectory of the effective federal funds rate (EFFR) in \autoref{fig:ffr.plot} and indicate the regimes. The first break is located in November 1994 after the EFFR sharply increases and the yield spread narrows, the second break is located in April 2003 during the recession. Here, the EFFR falls dramatically and we observe wider spreads. The third break is located in August 2010 after the Global Financial Crisis, and the fourth break is located in March 2015 at the end of the zero target rate regime. After the structural breaks are obtained, we re-estimate the model in each regime without scaling factors for higher precision and report the resulting coefficients in \autoref{tab:emp_term}. For comparison, the estimated breakpoints from the likelihood-based approach are located in January 1995, December 2003, July 2008, and July 2013 corresponding to roughly the same monetary policy events. The algorithm also detects four breakpoints with the sequential test clearly rejecting the hypothesis of three breaks in favor of a four break model. More than four breaks cannot be allocated if the default minimum regime length should be maintained (setting the trimming parameter to 0.15). 


\begin{table}[ht!]
\caption{Regime-specific coefficients of the term structure model}
\begin{center}
\scalebox{0.95}{
\begin{tabular}{c c c c c c}
\toprule[1pt]
 & \multicolumn{2}{c}{$r_{10y,t}$:} & & \multicolumn{2}{c}{$r_{5y,t}$:} \\
\cmidrule[0.5pt]{2-3}\cmidrule[0.5pt]{5-6}
Regimes: & $\hat{\mu}_1$ & $\hat{\beta}_1$ & & $\hat{\mu}_2$ & $\hat{\beta}_2$ \\
\midrule[0.5pt]
1990 m01 - 1994 m11 & 5.013 (0.261) & $0.464$ (0.082) & & 3.141 (0.154) & $0.678$ (0.055) \\
1994 m11 - 2003 m04 & 4.000 (0.211) & $0.392$ (0.125) & & 2.297 (0.163) & $0.653$ (0.093) \\
2003 m04 - 2010 m08 & 3.569 (0.329) & $0.267$ (0.109) & & 2.020 (0.220) & $0.556$ (0.077) \\
2010 m08 - 2015 m03 & 2.311 (0.259) & $-0.158$ (0.812) & & 1.103 (0.162) & $0.736$ (0.532) \\
2015 m03 - 2021 m07 & 1.220 (0.235) & $0.669$ (0.168) & & 0.525 (0.191) & $0.842$ (0.131) \\
\bottomrule[1pt]
\end{tabular}
}
\end{center}
\label{tab:emp_term}
\begin{tablenotes}
\scriptsize
\item Note: The coefficient estimates are obtained from a post-LASSO dynamic OLS estimation. We compute bootstrap standard errors based on 600 replications of the sieve bootstrap method for cointegrating regressions proposed in \cite{Chang2006}. Standard errors are computed for each identified regime separately and therefore do not take the estimation uncertainty regarding the structural break detection into account. Standard errors are given in parentheses.
\end{tablenotes}
\end{table}

It can be observed that the pairwise cointegrating vectors for most regimes substantially differ from $(1,-1)$. A simple $t$-test of the hypothesis does not lead to a rejection of the EHT for the last two regimes.\footnote{Note that the reported standard errors do not take the break estimation uncertainty into account.} While the estimates of the proportionality coefficient for the first three regimes in case of the 10-year maturity are reasonable (albeit still being able to reject the EHT), the estimated coefficient for the fourth regime from August 2010 to March 2015 is negative with a much larger standard error. The non-rejection of the EHT in this regime can therefore be attributed to the higher error term variance in this period. This regime can also be associated with a zero target rate and an unusually steep yield curve which might explain the unusually small proportionality coefficient for the 10-year maturity. The final regime is marked by relatively narrow yield spreads that begin to widen during the COVID-19 pandemic. Still, our results suggest that this most recent regime comes closest to satisfying the EHT. In summary, accounting for multiple structural breaks in the term structure model reveals some important differences for subsamples of the data but does not solve the term structure puzzle. Instead, it seems that the EHT does not hold for most of the sampling period.

\section{Conclusion}\label{sec:conc}
\noindent
We have proposed a computationally efficient alternative to the existing likelihood-based approach solving the change-point problem in multivariate systems with a mix of integrated and stationary regressors. Our two-step estimator is able to consistently determine the number of breaks, their timing and to jointly estimate the coefficients for each regime. This can be achieved without the need to conduct sequential tests and does not require generating critical values beyond those supplied in the original paper, for example, in situations with larger systems composed of many regressors. The algorithm is much faster than the dynamic programming algorithm used in the likelihood-approach and can solve change-point problems for large multivariate systems and several thousand observations in seconds. In turn, the likelihood-based approach allows for a straightforward construction of valid confidence bands by inverting the likelihood ratio test for a given break date \citep{EoMorley2015}. It remains to be investigated if such confidence bands can be constructed within the model selection approach taken in this paper. 

The crucial first step estimation is based on the group LASSO estimator and we utilize a group LARS algorithm to solve the change-point problem. Alternative choices of other penalties in the objective function could be used to potentially improve the estimator in some directions. For example, \cite{SafikhaniShojaie2020} use a fused LASSO penalty to allow the number of equations to grow with the sample size. Such high-dimensional extensions might be useful in panel settings where both the time and the cross-sectional dimension increase asymptotically. In principle, the proposed two-step estimator can also be applied to change-point problems in other important model classes. If we relaxed the strict exogeneity assumptions and instead used more restrictive assumptions about the error terms, for example, assuming Gaussian white noise errors, we could also deal with structural breaks in VECMs which involve a mix of integrated and stationary variables. Since this application poses additional technical difficulties, we leave this topic for future research.

\section*{Acknowledgements}
\noindent
The author thanks Konstantin Kuck, Thomas Dimpfl, Markus M\"o\ss{}ler, Robert Jung, and participants of the Research Seminar in Economics at the University of Hohenheim, the Asian Meeting of the Econometric Society in China 2022, the Econometric Society 2022 Australasia Meeting, the AMES 2022 Tokyo, and the EEA-ESEM in Milan for valuable comments. Further, he thanks Chun Yip Yau for sharing the programs for the group LARS algorithm, and Zhongjun Qu, Pierre Perron and Tatsushi Oka for sharing the programs for the likelihood-based approach. Funding by the German Research Foundation (Grant SCHW 2062/1-1) is gratefully acknowledged.

\clearpage

\appendix

\section{Mathematical Appendix}\label{sec:appendix}

\renewcommand{\theequation}{\thesection.\arabic{equation}}
\setcounter{equation}{0}


\begin{lemma}\label{lemma:1}
Under Assumption \ref{as:1}, we note that the following properties
\begin{eqnarray*}
&a)& \qquad \frac{1}{T} \sum\limits_{t=k}^{T} \frac{X_{l,t}}{\sqrt{T}} u_{j,t} = O_p(T^{-\frac{1}{2}}) \\
&b)& \qquad \frac{1}{T} \sum\limits_{t=k}^{T} \frac{t}{T} u_{j,t} = O_p(T^{-\frac{1}{2}}) \\
&c)& \qquad \frac{1}{T} \sum\limits_{t=k}^{T} u_{j,t} = O_p(T^{-\frac{1}{2}}) \\
&d)& \qquad \frac{1}{T} \sum\limits_{t=k}^{T} w_{l,t} u_{j,t} = O_p(T^{-\frac{1}{2}}),
\label{eq:lemma1}
\end{eqnarray*}
hold for $1 \leq j \leq q$ and all $k = 1, \dots, T$.
\end{lemma}

\begin{proofoflemma}
Property a) involving integrated regressors follows from Lemma A.4 in \cite{LiPerron2017}. The remaining properties are standard results for stationary processes.
\end{proofoflemma}
$\hfill \Box$


\begin{lemma}\label{lemma:2}
Under Assumption \ref{as:1}, for any $c_0 > 0$, there exists some constant $C > 0$ such that
\begin{equation}
P \left( \underset{1 \leq k \leq T}{\max} \, \underset{1 \leq l \leq d}{\max} \left\vert \frac{\boldsymbol{Z}(k,l)'\boldsymbol{U}}{T} \right\vert \geq c_0 \left( \frac{\log T}{T} \right)^{\frac{1}{4}} \right) \leq \frac{C}{c_0^4 \log T}.
\label{eq:lemma2}
\end{equation}
\end{lemma}

\begin{proofoflemma}
We first need to define some notation for the proof of this lemma: $\boldsymbol{Z}(k,.)$ is the $k$-th block column of $\boldsymbol{Z}$ and is of dimension $Tq \times d$. $\boldsymbol{Z}(k,l)$ is the $l$-th column of the $k$-th block column. Then, note that the following cases
\begin{eqnarray}
 &a)& \frac{\boldsymbol{Z}(k,l)'\boldsymbol{U}}{T} = \frac{1}{T} \sum\limits_{t=k}^{T} \frac{X_{i,t}}{\sqrt{T}} u_{j,t}, \qquad 1 \leq i \leq r, \quad 1 \leq j \leq q, \\
 &b)& \frac{\boldsymbol{Z}(k,l)'\boldsymbol{U}}{T} = \frac{1}{T} \sum\limits_{t=k}^{T} \frac{t}{T} u_{j,t}, \qquad 1 \leq j \leq q, \\
 &c)& \frac{\boldsymbol{Z}(k,l)'\boldsymbol{U}}{T} = \frac{1}{T} \sum\limits_{t=k}^{T} u_{j,t}, \qquad 1 \leq j \leq q, \\
 &d)& \frac{\boldsymbol{Z}(k,l)'\boldsymbol{U}}{T} = \frac{1}{T} \sum\limits_{t=k}^{T} w_{i,t} u_{j,t}, \qquad 1 \leq i \leq s, \quad 1 \leq j \leq q,
\end{eqnarray}
can appear in the full model depending on the choice of $l$. In each case, we obtain scalar partial sums. Although $\boldsymbol{Z}(k,l)$ is of dimension $Tq \times 1$, its structure guarantees that not more than $T$ elements are nonzero. 

We first turn to case a) and prove the lemma for this case in detail. Under Assumption \ref{as:1}, we have $E \left( \left\vert \frac{1}{T} \sum\limits_{t=1}^{k} \frac{X_{i,t}}{\sqrt{T}} u_{j,t} \right\vert^4 \right) \leq \frac{C}{32T^2}$ for $1 \leq i \leq r$, $1 \leq j \leq q$, $1 \leq k \leq T$, and all $T$. It follows that
\begin{eqnarray}
\notag P \left( \underset{1 \leq k \leq T}{\max} \left\vert \frac{1}{T} \sum\limits_{t=1}^{k} \frac{X_{i,t}}{\sqrt{T}} u_{j,t} \right\vert \geq \frac{c_0}{2} \left( \frac{\log T}{T} \right)^{\frac{1}{4}} \right) &\leq& \sum\limits_{k=1}^{T} P \left( \left\vert  \frac{1}{T} \sum\limits_{t=1}^{k} \frac{X_{i,t}}{\sqrt{T}} u_{j,t} \right\vert \geq \frac{c_0}{2} \left( \frac{\log T}{T} \right)^{\frac{1}{4}} \right) \\
 &\leq& \sum\limits_{k=1}^{T} \frac{16 T}{c_0^4 \log T} E\left( \left\vert \frac{1}{T} \sum\limits_{t=1}^{k} \frac{X_{i,t}}{\sqrt{T}} u_{j,t} \right\vert^4 \right)\\
\notag &\leq& \frac{C}{2c_0^4 \log T},
\end{eqnarray}
for all $i,j$, and some $C > 0$. Thus, it holds that
\begin{eqnarray}
\notag && P \left( \underset{1 \leq k \leq T}{\max} \left\vert \frac{1}{T} \sum\limits_{t=k}^{T} \frac{X_{i,t}}{\sqrt{T}} u_{j,t} \right\vert \geq c_0 \left( \frac{\log T}{T} \right)^{\frac{1}{4}} \right) \\
&=& P \left( \underset{1 \leq k \leq T}{\max} \left\vert \frac{1}{T} \sum\limits_{t=1}^{T} \frac{X_{i,t}}{\sqrt{T}} u_{j,t} - \frac{1}{T}  \sum\limits_{i=1}^{k-1} \frac{X_{l,t}}{\sqrt{T}} u_{j,t} \right\vert \geq c_0 \left( \frac{\log T}{T} \right)^{\frac{1}{4}} \right) \\
\notag &\leq& P \left( \left\vert \frac{1}{T}  \sum\limits_{t=1}^{T} \frac{X_{i,t}}{\sqrt{T}} u_{j,t} \right\vert \geq \frac{c_0}{2} \left( \frac{\log T}{T} \right)^{\frac{1}{4}} \right) + P \left( \underset{1 \leq k \leq T}{\max} \left\vert \frac{1}{T} \sum\limits_{t=1}^{k-1} \frac{X_{i,t}}{\sqrt{T}} u_{j,t} \right\vert \geq \frac{c_0}{2} \left( \frac{\log T}{T} \right)^{\frac{1}{4}} \right) \\
\notag &\leq& \frac{C}{c_0^4 \log T}.
\end{eqnarray}

Now, considering that all components of $d = q(r + 2 + s)$ are finite, Equation~\eqref{eq:lemma2} follows for case a). The remaining cases are proven similarly, relying on the moment bounds
\begin{eqnarray}
E\left( \left\vert \frac{1}{T}  \sum\limits_{t=1}^{k} \frac{t}{T} u_{j,t} \right\vert^4 \right) 
&\leq& \frac{1}{T^2} E\left( \left\vert \sum\limits_{t=1}^{k} \frac{u_{j,t}}{\sqrt{T}} \right\vert^4 \right) \leq \frac{C}{32T^2}, \qquad 1 \leq j \leq q,
\end{eqnarray}
and
\begin{equation}
E\left( \left\vert \frac{1}{T} \sum\limits_{t=1}^{k} w_{i,t} u_{j,t} \right\vert^4 \right) = \frac{1}{T^2} E\left( \left\vert \sum\limits_{t=1}^{k} w_{i,t} u_{j,t} \right\vert^4 \right) \leq \frac{C}{32T^2}, \qquad 1 \leq i \leq s, \quad 1 \leq j \leq q,
\end{equation}
respectively. The last inequality in both cases holds, because the processes $u_t$ and $w_t$ are stationary and the partial sums scaled by $1/\sqrt{T}$ have finite fourth moments for each $T$ according to Assumption \ref{as:1}.
\end{proofoflemma}
$\hfill \Box$



\begin{lemma}\label{lemma:3}
Let $\hat{\boldsymbol{\theta}}(T)$ be the estimator of $\boldsymbol{\theta}(T)$ as defined in Equation~\eqref{eq:gfl2} and $\bar{Z}_t = (Z_t' \otimes I)$, then it holds under the same conditions as in Theorem \ref{th:1} that
\begin{equation*}
\sum\limits_{s = \hat{t}_j}^{T} \bar{Z}_s' \left( Y_s - \bar{Z}_s \sum\limits_{i=1}^{s} \boldsymbol{K}' \hat{\boldsymbol{\theta}}_i \right) - \frac{1}{2} T \lambda_T \frac{\boldsymbol{K}' \hat{\boldsymbol{\theta}}_{\hat{t}_j}}{\Vert \boldsymbol{K}' \hat{\boldsymbol{\theta}}_{\hat{t}_j} \Vert} = \boldsymbol{0}, \qquad \forall \hat{\boldsymbol{\theta}}_{\hat{t}_j} \neq \boldsymbol{0},
\end{equation*}
and 
\begin{equation*}
\left\Vert \sum\limits_{s = j}^{T} \bar{Z}_s' \left( Y_s - \bar{Z}_s \sum\limits_{i=1}^{s} \boldsymbol{K}' \hat{\boldsymbol{\theta}}_i \right) \right\Vert \leq \frac{1}{2} T \lambda_T, \qquad \forall j.
\end{equation*}
\end{lemma}

\begin{proofoflemma}
This lemma is a direct consequence of the Karush-Kuhn-Tucker (KKT) conditions for group LASSO estimators.
\end{proofoflemma}
$\hfill \Box$


\begin{lemma}\label{lemma:5}
Under Assumptions \ref{as:1} - \ref{as:5}, if $m_0$ is known and $|\mathcal{A}_T| = m_0$, then
\begin{equation*}
P \left( \underset{1 \leq j \leq m_0}{\max} | \hat{t}_j - t^0_j | \leq T \gamma_T \right) \to 1, \qquad \text{as } T \to \infty.
\end{equation*}
\end{lemma}

\begin{proofoflemma}
Define $A_{Ti} = \left\lbrace | \hat{t}_i - t^0_i | > T \gamma_T \right\rbrace$, $i = 1,2, \dots, m_0$ such that
\begin{equation}
P \left( \underset{1 \leq i \leq m_0}{\max} | \hat{t}_i - t^0_i | > T \gamma_T \right) \leq \sum\limits_{i=1}^{m_0} P \left( | \hat{t}_i - t^0_i | > T \gamma_T \right) = \sum\limits_{i=1}^{m_0} P \left( A_{Ti} \right).
\end{equation}
Further define $C_T = \left\lbrace \max_{1 \leq i \leq m_0} | \hat{t}_i - t^0_i | \leq \min_i \, | t^0_i - t^0_{i-1} |/2 \right\rbrace$. It suffices to show that
\begin{equation}
\sum\limits_{i=1}^{m_0} P \left( A_{Ti} C_T \right) \to 0 \text{ and } \sum\limits_{i=1}^{m_0} P \left( A_{Ti} C^c_T \right) \to 0.
\end{equation}
The proof follows along the lines of the proof of Theorem 2.2 in \cite{ChanYauZhang2014} but instead of Lemma A.2 in their paper which bounds the tail probability for $\beta$-mixing stationary processes, we rely on the moment conditions stated in Assumptions \ref{as:1} and the strict exogeneity condition in Assumption \ref{as:2} to do so. In the following, we focus on $\sum_{i=1}^{m_0} P \left( A_{Ti} C_T \right) \to 0$ because the complementary part can be shown using similar arguments.

In the set $C_T$, it holds that 
\begin{equation}
t^0_{i-1} < \hat{t}_i < t^0_{i+1}, \qquad \forall \, 1 \leq i \leq m_0. 
\end{equation}
Next, we split $A_{Ti}$ into two parts (i) $\hat{t}_i < t^0_i$ and (ii) $\hat{t}_i > t^0_i$ to show that $P \left( A_{Ti} C_T \right) \to 0$.

In case of (i), applying Lemma \ref{lemma:3} (KKT conditions) yields
\begin{equation}
\Vert \sum\limits_{l=\hat{t}_i}^{t^0_i-1} \bar{Z}_l' (Y_l - \bar{Z}_l \sum\limits_{j=1}^{\hat{t}_{i+1}-1} \boldsymbol{K}' \hat{\boldsymbol{\theta}}_j) \Vert \leq T \lambda_T.
\end{equation}
Note that because of $\hat{t}_i < t^0_i$, the true coefficient has not yet changed at $\hat{t}_i$. Hence, plugging in for $Y_l = \bar{Z}_l \sum_{i=j}^{t^0_i-1} \boldsymbol{K}' \boldsymbol{\theta}^0_j + u_l$ yields
\begin{equation}
\Vert \sum\limits_{l=\hat{t}_i}^{t^0_i-1} \bar{Z}_l' u_l + \sum\limits_{l=\hat{t}_i}^{t^0_i-1} \bar{Z}_l' \bar{Z}_l \left(\sum\limits_{j=1}^{t^0_i-1} \boldsymbol{K}' \boldsymbol{\theta}^0_j - \sum\limits_{j=1}^{t^0_{i+1}-1} \boldsymbol{K}' \boldsymbol{\theta}^0_j \right) + \sum\limits_{l=\hat{t}_i}^{t^0_i-1} \bar{Z}_l' \bar{Z}_l \left(\sum\limits_{j=1}^{t^0_{i+1}-1} \boldsymbol{K}' \boldsymbol{\theta}^0_j - \sum\limits_{j=1}^{\hat{t}_{i+1}-1} \boldsymbol{K}' \hat{\boldsymbol{\theta}}_j \right) \Vert \leq T \lambda_T.
\end{equation}
It follows for $\hat{t}_i < t^0_i$ that,
\begin{eqnarray}
\notag & P \left( A_{Ti} C_T \right) \leq P \left( \left\lbrace \frac{1}{3} \Vert \sum\limits_{l=\hat{t}_i}^{t^0_i-1} \bar{Z}_l' \bar{Z}_l \left(\sum\limits_{j=1}^{t^0_i-1} \boldsymbol{K}' \boldsymbol{\theta}^0_j - \sum\limits_{j=1}^{t^0_{i+1}-1} \boldsymbol{K}' \boldsymbol{\theta}^0_j \right) \Vert \leq T \lambda_T \right\rbrace \cap \left\lbrace | \hat{t}_i - t^0_i | > T \gamma_T \right\rbrace \right) \\
\notag &+ P \left( \left\lbrace \Vert \sum\limits_{l=\hat{t}_i}^{t^0_i-1} \bar{Z}_l' u_l \Vert > \frac{1}{3} \Vert \sum\limits_{l=\hat{t}_i}^{t^0_i-1} \bar{Z}_l' \bar{Z}_l \left(\sum\limits_{j=1}^{t^0_i-1} \boldsymbol{K}' \boldsymbol{\theta}^0_j - \sum\limits_{j=1}^{t^0_{i+1}-1} \boldsymbol{K}' \boldsymbol{\theta}^0_j \right) \Vert \right\rbrace \cap \left\lbrace | \hat{t}_i - t^0_i | > T \gamma_T \right\rbrace \right) \\
\notag &+ P \left( \left\lbrace \Vert \sum\limits_{l=\hat{t}_i}^{t^0_i-1} \bar{Z}_l' \bar{Z}_l \left( \sum\limits_{j=1}^{t^0_{i+1}-1} \boldsymbol{K}' \boldsymbol{\theta}^0_j - \sum\limits_{j=1}^{\hat{t}_{i+1}-1} \boldsymbol{K}' \hat{\boldsymbol{\theta}}_j \right) \Vert \right. \right. \\
& \left. \left. > \frac{1}{3} \Vert \sum\limits_{l=\hat{t}_i}^{t^0_i-1} \bar{Z}_l' \bar{Z}_l \left(\sum\limits_{j=1}^{t^0_i-1} \boldsymbol{K}' \boldsymbol{\theta}^0_j - \sum\limits_{j=1}^{t^0_{i+1}-1} \boldsymbol{K}' \boldsymbol{\theta}^0_j \right) \Vert \right\rbrace \cap A_{Ti}C_T \right) \\
\notag &= P \left( A_{Ti1} \right) + P \left( A_{Ti2} \right) + P \left( A_{Ti3} \right).
\end{eqnarray}


For the first term, it can be shown under Assumptions \ref{as:1} and \ref{as:2} and on the set $\left\lbrace | \hat{t}_i - t^0_i | > T \gamma_T \right\rbrace$ that
\begin{eqnarray}
&& \frac{1}{3} \Vert \sum\limits_{l=\hat{t}_i}^{t^0_i-1} \bar{Z}_l' \bar{Z}_l \left(\sum\limits_{j=1}^{t^0_i-1} \boldsymbol{K}' \boldsymbol{\theta}^0_j - \sum\limits_{j=1}^{t^0_{i+1}-1} \boldsymbol{K}' \boldsymbol{\theta}^0_j \right) \Vert \\
&\geq& \frac{\vert \hat{t}_i - t^0_i \vert}{6} \Vert E (\bar{Z}_{t^0_i-1}' \bar{Z}_{t^0_i-1}) \left(\sum\limits_{j=1}^{t^0_i-1} \boldsymbol{K}' \boldsymbol{\theta}^0_j - \sum\limits_{j=1}^{t^0_{i+1}-1} \boldsymbol{K}' \boldsymbol{\theta}^0_j \right) \Vert \\
&\geq& \frac{T \gamma_T}{6} \Vert E (\bar{Z}_{t^0_i-1}' \bar{Z}_{t^0_i-1}) \left(\sum\limits_{j=1}^{t^0_i-1} \boldsymbol{K}' \boldsymbol{\theta}^0_j - \sum\limits_{j=1}^{t^0_{i+1}-1} \boldsymbol{K}' \boldsymbol{\theta}^0_j \right) \Vert = c_0 T \gamma_T,
\label{eq:AT1}
\end{eqnarray}
holds with probability going to one. Taking into account that $\gamma_T/\lambda_T \to \infty$, we conclude that $P \left( A_{Ti1} \right) \to 0$ for $T \to \infty$.
For the second term, we have
\begin{equation}
\Vert \sum\limits_{l=\hat{t}_i}^{t^0_i-1} \bar{Z}_l' u_l \Vert = O_p((T\gamma_T)^{\frac{1}{2}}),
\end{equation}
using Lemma \ref{lemma:2} but since $\frac{1}{3} \Vert \sum_{l=\hat{t}_i}^{t^0_i-1} \bar{Z}_l' \bar{Z}_l \left(\sum_{j=1}^{t^0_i-1} \boldsymbol{K}' \boldsymbol{\theta}^0_j - \sum_{j=1}^{t^0_{i+1}-1} \boldsymbol{K}' \boldsymbol{\theta}^0_j \right) \Vert > c_0 T \gamma_T$ with probability going to one, we conclude that the right hand side of the inequality asymptotically dominates the left hand side and $P \left( A_{Ti2} \right) \to 0$ for $T \to \infty$. 

Turning to the third term, we apply Lemma \ref{lemma:3} to the interval $[t^0_i, (t^0_i + t^0_{i+1})/2]$ which yields
\begin{equation}
    \Vert \sum\limits_{l=t^0_i}^{(t^0_i + t^0_{i+1})/2 - 1} \bar{Z}_l' \bar{Z}_l \left( \sum\limits_{j=1}^{t^0_{i+1}-1} \boldsymbol{K}' \boldsymbol{\theta}^0_j - \sum\limits_{j=1}^{\hat{t}_{i+1}-1} \boldsymbol{K}' \hat{\boldsymbol{\theta}}_j \right) \Vert \leq T\lambda_T + \Vert \sum\limits_{l=t^0_i}^{(t^0_i + t^0_{i+1})/2 - 1} \bar{Z}_l' u_l \Vert.
    \label{eq:ATi3}
\end{equation}
Since $|t^0_{i+1} - t^0_i| \geq 4T\gamma_T$, it follows that for any $x > 0$,
\begin{equation}
    \Vert \sum\limits_{l=t^0_i}^{(t^0_i + t^0_{i+1})/2 - 1} \bar{Z}_l' u_l \Vert \leq x |t^0_{i+1} - t^0_i|.
\end{equation}
However, it also holds that
\begin{eqnarray}
    &&\Vert \sum\limits_{l=t^0_i}^{(t^0_i + t^0_{i+1})/2 - 1} \bar{Z}_l' \bar{Z}_l \left( \sum\limits_{j=1}^{t^0_{i+1}-1} \boldsymbol{K}' \boldsymbol{\theta}^0_j - \sum\limits_{j=1}^{\hat{t}_{i+1}-1} \boldsymbol{K}' \hat{\boldsymbol{\theta}}_j \right) \Vert \\
    &\geq& \frac{\vert t^0_{i+1} - t^0_i \vert}{4} \Vert E (\bar{Z}_{t^0_{i+1}-1}' \bar{Z}_{t^0_{i+1}-1}) \left(\sum\limits_{j=1}^{t^0_{i+1}-1} \boldsymbol{K}' \boldsymbol{\theta}^0_j - \sum\limits_{j=1}^{\hat{t}_{i+1}-1} \boldsymbol{K}' \hat{\boldsymbol{\theta}}_j \right) \Vert,
\end{eqnarray}
with probability approaching one. Combining both results yields
\begin{eqnarray}
    \Vert E (\bar{Z}_{t^0_i-1}' \bar{Z}_{t^0_i-1}) \left(\sum\limits_{j=1}^{t^0_{i+1}-1} \boldsymbol{K}' \boldsymbol{\theta}^0_j - \sum\limits_{j=1}^{\hat{t}_{i+1}-1} \boldsymbol{K}' \hat{\boldsymbol{\theta}}_j \right) \Vert \leq \frac{4 T \lambda_T}{\vert t^0_{i+1} - t^0_i \vert} + 4x.
\end{eqnarray}
It follows with probability approaching one that
\begin{eqnarray}
    && \Vert \sum\limits_{l=\hat{t}_i}^{t^0_i-1} \bar{Z}_l' \bar{Z}_l \left( \sum\limits_{j=1}^{t^0_{i+1}-1} \boldsymbol{K}' \boldsymbol{\theta}^0_j - \sum\limits_{j=1}^{\hat{t}_{i+1}-1} \boldsymbol{K}' \hat{\boldsymbol{\theta}}_j \right) \Vert \\
    &\leq& 2(t^0_i - \hat{t}_i) \Vert E (\bar{Z}_{t^0_i-1}' \bar{Z}_{t^0_i-1}) \left(\sum\limits_{j=1}^{t^0_{i+1}-1} \boldsymbol{K}' \boldsymbol{\theta}^0_j - \sum\limits_{j=1}^{\hat{t}_{i+1}-1} \boldsymbol{K}' \hat{\boldsymbol{\theta}}_j \right) \Vert \\
    &\leq& \frac{C_1 T \lambda_T (t^0_i - \hat{t}_i)}{(t^0_{i+1} - t^0_i)} + C_2 x (t^0_i - \hat{t}_i),
\end{eqnarray}
but at the same time $\frac{1}{3} \Vert \sum_{l=\hat{t}_i}^{t^0_i-1} \bar{Z}_l' \bar{Z}_l \left(\sum_{j=1}^{t^0_i-1} \boldsymbol{K}' \boldsymbol{\theta}^0_j - \sum_{j=1}^{t^0_{i+1}-1} \boldsymbol{K}' \boldsymbol{\theta}^0_j \right) \Vert > c_0 (t^0_i - \hat{t}_i)$ with probability going to one according to Equation~\eqref{eq:AT1}. Since $T\gamma_T/(t^0_{i+1} - t^0_i) \to 0$ according to Assumption \ref{as:5}, this implies that $P \left( A_{Ti3} \right) \to 0$ for $T \to \infty$. It follows that $P \left( A_{Ti} C_T \cap \lbrace \hat{t}_i < t^0_i \rbrace \right) \to 0$.

In case of (ii), we have
\begin{equation}
\Vert \sum\limits_{l=t^0_i}^{\hat{t}_i-1} \bar{Z}_l' (Y_l - \bar{Z}_l \sum\limits_{j=1}^{\hat{t}_i-1} \boldsymbol{K}' \hat{\boldsymbol{\theta}}_j) \Vert \leq T \lambda_T,
\end{equation}

Since $\hat{t}_i > t^0_i$, the true coefficient has changed at $t^0_i$ and we plug in for $Y_l = \bar{Z}_l' \sum_{j=1}^{t^0_{i+1}-1} \boldsymbol{K}' \boldsymbol{\theta}^0_j + u_l$, which yields
\begin{equation}
\Vert \sum\limits_{l=t^0_i}^{\hat{t}_i-1} \bar{Z}_l' u_l + \sum\limits_{l=t^0_i}^{\hat{t}_i-1} \bar{Z}_l' \bar{Z}_l \left(\sum\limits_{j=1}^{t^0_{i+1}-1} \boldsymbol{K}' \boldsymbol{\theta}^0_j - \sum\limits_{j=1}^{t^0_i-1} \boldsymbol{K}' \boldsymbol{\theta}^0_j \right) + \sum\limits_{l=t^0_i}^{\hat{t}_i-1} \bar{Z}_l' \bar{Z}_l \left(\sum\limits_{j=1}^{t^0_i-1} \boldsymbol{K}' \boldsymbol{\theta}^0_j - \sum\limits_{j=1}^{\hat{t}_i-1} \boldsymbol{K}' \hat{\boldsymbol{\theta}}_j \right) \Vert \leq T \lambda_T.
\end{equation}
It follows that,
\begin{eqnarray}
\notag & P \left( A_{Ti} C_T \right) \leq P \left( \left\lbrace \frac{1}{3} \Vert \sum\limits_{l=t^0_i}^{\hat{t}_i-1} \bar{Z}_l' \bar{Z}_l \left(\sum\limits_{j=1}^{t^0_{i+1}-1} \boldsymbol{K}' \boldsymbol{\theta}^0_j - \sum\limits_{j=1}^{t^0_i-1} \boldsymbol{K}' \boldsymbol{\theta}^0_j \right) \Vert \leq T \lambda_T \right\rbrace \cap \left\lbrace | \hat{t}_i - t^0_i | > T \gamma_T \right\rbrace \right) \\
\notag &+ P \left( \left\lbrace \Vert \sum\limits_{l=t^0_i}^{\hat{t}_i-1} \bar{Z}_l' u_l \Vert > \frac{1}{3} \Vert \sum\limits_{l=t^0_i}^{\hat{t}_i-1} \bar{Z}_l' \bar{Z}_l \left(\sum\limits_{j=1}^{t^0_{i+1}-1} \boldsymbol{K}' \boldsymbol{\theta}^0_j - \sum\limits_{j=1}^{t^0_i-1} \boldsymbol{K}' \boldsymbol{\theta}^0_j \right) \Vert \right\rbrace \cap \left\lbrace | \hat{t}_i - t^0_i | > T \gamma_T \right\rbrace \right) \\
\notag &+ P \left( \left\lbrace \Vert \sum\limits_{l=t^0_i}^{\hat{t}_i-1} \bar{Z}_l' \bar{Z}_l \left(\sum\limits_{j=1}^{t^0_i-1} \boldsymbol{K}' \boldsymbol{\theta}^0_j - \sum\limits_{j=1}^{\hat{t}_i-1} \boldsymbol{K}' \hat{\boldsymbol{\theta}}_j \right) \Vert \right. \right. \\
& \left. \left. > \frac{1}{3} \Vert \sum\limits_{l=t^0_i}^{\hat{t}_i-1} \bar{Z}_l' \bar{Z}_l \left(\sum\limits_{j=1}^{t^0_{i+1}-1} \boldsymbol{K}' \boldsymbol{\theta}^0_j - \sum\limits_{j=1}^{t^0_i-1} \boldsymbol{K}' \boldsymbol{\theta}^0_j \right) \Vert \right\rbrace \cap A_{Ti}C_T \right) \\
\notag &= P \left( A_{Ti1} \right) + P \left( A_{Ti2} \right) + P \left( A_{Ti3} \right).
\end{eqnarray}
The same arguments as for case (i) can be used to show that $P \left( A_{Ti} C_T \cap \lbrace \hat{t}_i > t^0_i \rbrace \right) \to 0$. Combining (i) and (ii) completes the proof of $P \left( A_{Ti} C_T \right) \to 0$.
\end{proofoflemma}
$\hfill \Box$


\begin{proofof}
By the definition of the group LASSO estimator, we obtain
\begin{eqnarray}
\frac{1}{T} \left\Vert \boldsymbol{Y} - \boldsymbol{Z} \hat{\boldsymbol{\theta}}(T) \right\Vert^2 + \lambda_T \sum\limits_{i=1}^{T} \Vert \hat{\boldsymbol{\theta}}_i \Vert \leq \frac{1}{T} \left\Vert \boldsymbol{Y} - \boldsymbol{Z} \boldsymbol{\theta}^0(T) \right\Vert^2 + \lambda_T \sum\limits_{i=1}^{T} \Vert \boldsymbol{\theta}^0_i \Vert.
\label{eq:gl.ineq}
\end{eqnarray}
Denoting $\bar{\mathcal{A}} = \lbrace t_0, t_1, t_2, \dots, t_{m_0} \rbrace$ and inserting $\boldsymbol{Y} = \boldsymbol{Z} \boldsymbol{\theta}^0(T) + \boldsymbol{U}$ into Inequality~\eqref{eq:gl.ineq}, we have
\begin{eqnarray}
\frac{1}{T} \Vert \boldsymbol{Z} (\boldsymbol{\theta}^0(T) - \hat{\boldsymbol{\theta}}(T)) \Vert^2 &\leq& \frac{2}{T} (\boldsymbol{\theta}^0(T) - \hat{\boldsymbol{\theta}}(T))' \boldsymbol{Z}' \boldsymbol{U} + \lambda_T \sum\limits_{i=1}^{T} \Vert \boldsymbol{\theta}^0_i \Vert - \lambda_T \sum\limits_{i=1}^{T} \Vert \hat{\boldsymbol{\theta}}_i \Vert \\
\notag &\leq& 2 d \left\Vert \frac{\boldsymbol{Z}'\boldsymbol{U}}{T} \right\Vert_{\infty} \left( \sum\limits_{i=1}^{T} \Vert \boldsymbol{\theta}^0_i - \hat{\boldsymbol{\theta}_i} \Vert \right) + \lambda_T \sum\limits_{i=1}^{T} \Vert \boldsymbol{\theta}^0_i \Vert - \lambda_T \sum\limits_{i=1}^{T} \Vert \hat{\boldsymbol{\theta}}_i \Vert \\
\notag &\leq& 2 d c_0 \left( \frac{\log T}{T} \right)^{\frac{1}{4}} \left( \sum\limits_{i=1}^{T} \Vert \boldsymbol{\theta}^0_i - \hat{\boldsymbol{\theta}_i} \Vert \right) + \lambda_T \sum\limits_{i \in \bar{\mathcal{A}}}^{} (\Vert \boldsymbol{\theta}^0_i \Vert - \Vert \hat{\boldsymbol{\theta}}_i \Vert) - \lambda_T \sum\limits_{i \in \bar{\mathcal{A}}^c}^{} \Vert \hat{\boldsymbol{\theta}}_i \Vert \\
\notag &\leq& \lambda_T \sum\limits_{i \in \bar{\mathcal{A}}}^{} \Vert \boldsymbol{\theta}^0_i - \hat{\boldsymbol{\theta}_i} \Vert + \lambda_T \sum\limits_{i \in \bar{\mathcal{A}}}^{} (\Vert \boldsymbol{\theta}^0_i \Vert - \Vert \hat{\boldsymbol{\theta}}_i \Vert) \\
\notag &\leq& 2 \lambda_T \sum\limits_{i \in \bar{\mathcal{A}}}^{} \Vert \boldsymbol{\theta}^0_i \Vert \leq 2 \lambda_T (m_0 + 1) \max_{1 \leq j \leq m_0 + 1} \Vert \boldsymbol{\theta}^0_{t^0_{j-1}} \Vert \\
\notag &=& 4 d c_0 \left( \frac{\log T}{T} \right)^{\frac{1}{4}} (m_0 + 1) M_{\theta}.
\end{eqnarray}
with probability greater than $1 - \frac{C}{c_0^4 \log T}$ according to Lemma \ref{lemma:2}.
\end{proofof}
$\hfill \Box$


\begin{proofof}
We begin to prove the first part. Suppose that $|\mathcal{A}_T| < m_0$, then there exists some $t^0_{i_0}$, $i_0 = 1,2, \dots$ and $\hat{t}_{l_0} \in \mathcal{A}_T \cup \lbrace 0, \infty \rbrace$, $l_0 = 0,1, \dots, |\mathcal{A}_T| + 1$ with $t^0_{i_0+1} - t^0_{i_0} \vee \hat{t}_{l_0} \geq T \gamma_T/3$ and $t^0_{i_0+2} \wedge \hat{t}_{l_0+1} - t^0_{i_0+1} \geq T\gamma_T/3$ where $\hat{t}_0 = 0$ and $\hat{t}_{|\mathcal{A}_T| + 1} = \infty$.


First, applying Lemma \ref{lemma:3} to the interval $[t^0_{i_0} \vee \hat{t}_{l_0}, t^0_{i_0+1}-1]$ yields
\begin{equation}
\Vert \sum\limits_{l=t^0_{i_0} \vee \hat{t}_{l_0}}^{t^0_{i_0+1}-1} \bar{Z}_l' (Y_l - \bar{Z}_l \sum\limits_{j=1}^{\hat{t}_{l_0+1}-1} \boldsymbol{K}' \hat{\boldsymbol{\theta}}_j) \Vert \leq T \lambda_T.
\end{equation}
Note that the true coefficient has changed at $t^0_{i_0}$ but does not change until $t^0_{i_0+1}$. Hence, plugging in for $Y_l = \bar{Z}_l \sum_{j=1}^{t^0_{i_0+1}-1} \boldsymbol{K}' \boldsymbol{\theta}^0_j + u_l$ yields
\begin{eqnarray}
\Vert \sum\limits_{l=t^0_{i_0} \vee \hat{t}_{l_0}}^{t^0_{i_0+1}-1} \bar{Z}_l' u_l + \sum\limits_{l=t^0_{i_0} \vee \hat{t}_{l_0}}^{t^0_{i_0+1}-1} \bar{Z}_l' \bar{Z}_l \left(\sum\limits_{j=1}^{t^0_{i_0+1}-1} \boldsymbol{K}' \boldsymbol{\theta}^0_j - \sum\limits_{j=1}^{t^0_{i_0}-1} \boldsymbol{K}' \boldsymbol{\theta}^0_j \right) \\
\notag + \sum\limits_{l=t^0_{i_0} \vee \hat{t}_{l_0}}^{t^0_{i_0+1}-1} \bar{Z}_l' \bar{Z}_l \left(\sum\limits_{j=1}^{t^0_{i_0}-1} \boldsymbol{K}' \boldsymbol{\theta}^0_j - \sum\limits_{j=1}^{\hat{t}_{l_0+1}-1} \boldsymbol{K}' \hat{\boldsymbol{\theta}}_j \right) \Vert \leq T \lambda_T,
\end{eqnarray}
and 
\begin{equation}
\Vert \sum\limits_{l=t^0_{i_0} \vee \hat{t}_{l_0}}^{t^0_{i_0+1}-1} \bar{Z}_l' \bar{Z}_l \left(\sum\limits_{j=1}^{t^0_{i_0}-1} \boldsymbol{K}' \boldsymbol{\theta}^0_j - \sum\limits_{j=1}^{\hat{t}_{l_0+1}-1} \boldsymbol{K}' \hat{\boldsymbol{\theta}}_j \right) \Vert \leq T \lambda_T + \Vert \sum\limits_{l=t^0_{i_0} \vee \hat{t}_{l_0}}^{t^0_{i_0+1}-1} \bar{Z}_l' u_l \Vert.
\end{equation}

Second, applying Lemma \ref{lemma:3} to the interval $[t^0_{i_0+1}, t^0_{i_0+2} \wedge \hat{t}_{l_0+1}-1]$ yields
\begin{equation}
\Vert \sum\limits_{l=t^0_{i_0+1}}^{t^0_{i_0+2} \wedge \hat{t}_{l_0+1}-1} \bar{Z}_l' (Y_l - \bar{Z}_l \sum\limits_{j=1}^{\hat{t}_{l_0+1}-1} \boldsymbol{K}' \hat{\boldsymbol{\theta}}_j) \Vert \leq T \lambda_T.
\end{equation}
Since the true coefficient has changed at $t^0_{i_0+1}$ but does not change again until $t^0_{i_0+2}$, we plug in $Y_l = \bar{Z}_l \sum_{j=1}^{t^0_{i_0+2}-1} \boldsymbol{K}' \boldsymbol{\theta}^0_j + u_l$ which yields
\begin{eqnarray}
\Vert \sum\limits_{l=t^0_{i_0+1}}^{t^0_{i_0+2} \wedge \hat{t}_{l_0+1}-1} \bar{Z}_l' u_l + \sum\limits_{l=t^0_{i_0+1}}^{t^0_{i_0+2} \wedge \hat{t}_{l_0+1}-1} \bar{Z}_l' \bar{Z}_l \left(\sum\limits_{j=1}^{t^0_{i_0+2}-1} \boldsymbol{K}' \boldsymbol{\theta}^0_j - \sum\limits_{j=1}^{t^0_{i_0+1}-1} \boldsymbol{K}' \boldsymbol{\theta}^0_j \right) \\
\notag + \sum\limits_{l=t^0_{i_0+1}}^{t^0_{i_0+2} \wedge \hat{t}_{l_0+1}-1} \bar{Z}_l' \bar{Z}_l \left(\sum\limits_{j=1}^{t^0_{i_0+1}-1} \boldsymbol{K}' \boldsymbol{\theta}^0_j - \sum\limits_{j=1}^{\hat{t}_{l_0+1}-1} \boldsymbol{K}' \hat{\boldsymbol{\theta}}_j \right) \Vert \leq T \lambda_T,
\end{eqnarray}
and 
\begin{equation}
\Vert \sum\limits_{l=t^0_{i_0+1}}^{t^0_{i_0+2} \wedge \hat{t}_{l_0+1}-1} \bar{Z}_l' \bar{Z}_l \left(\sum\limits_{j=1}^{t^0_{i_0+1}-1} \boldsymbol{K}' \boldsymbol{\theta}^0_j - \sum\limits_{j=1}^{\hat{t}_{l_0+1}-1} \boldsymbol{K}' \hat{\boldsymbol{\theta}}_j \right) \Vert \leq T \lambda_T + \Vert \sum\limits_{l=t^0_{i_0+1}}^{t^0_{i_0+2} \wedge \hat{t}_{l_0+1}-1} \bar{Z}_l' u_l \Vert.
\end{equation}

As in the proof of Lemma \ref{lemma:5}, it can be shown that 
\begin{equation}
\Vert E( \bar{Z}_{t^0_{i_0+1}}' \bar{Z}_{t^0_{i_0+1}} ) \left( \sum\limits_{j=1}^{t^0_{i_0}-1} \boldsymbol{K}' \boldsymbol{\theta}^0_j - \sum\limits_{j=1}^{\hat{t}_{l_0+1}-1} \boldsymbol{K}' \hat{\boldsymbol{\theta}}_j \right) \Vert \leq 2 T \lambda_T / (t^0_{i_0+1} - t^0_{i_0} \vee \hat{t}_{l_0}) + 2x,
\label{eq:vee}
\end{equation}
and
\begin{equation}
\Vert E( \bar{Z}_{t^0_{i_0+2}}' \bar{Z}_{t^0_{i_0+2}} ) \left( \sum\limits_{j=1}^{t^0_{i_0+1}-1} \boldsymbol{K}' \boldsymbol{\theta}^0_j - \sum\limits_{j=1}^{\hat{t}_{l_0+1}-1} \boldsymbol{K}' \hat{\boldsymbol{\theta}}_j \right) \Vert \leq 2 T \lambda_T / (t^0_{i_0+2} \wedge \hat{t}_{l_0+1} - t^0_{i_0+1}) + 2x,
\label{eq:wedge}
\end{equation}
holds for arbitrary $x > 0$ with probability approaching one. Since $t^0_{i_0+1} - t^0_{i_0} \vee \hat{t}_{l_0} \geq T \gamma_T/3$, $t^0_{i_0+2} \wedge \hat{t}_{l_0+1} - t^0_{i_0+1} \geq T\gamma_T/3$, and $\gamma_T / \lambda_T \to \infty$, it follows that $\Vert \sum_{j=1}^{t^0_{i_0}-1} \boldsymbol{K}' \boldsymbol{\theta}^0_j - \sum_{j=1}^{\hat{t}_{l_0+1}-1} \boldsymbol{K}' \hat{\boldsymbol{\theta}}_j \Vert \overset{p}{\to} 0$ and $\Vert \sum_{j=1}^{t^0_{i_0+1}-1} \boldsymbol{K}' \boldsymbol{\theta}^0_j - \sum_{j=1}^{\hat{t}_{l_0+1}-1} \boldsymbol{K}' \hat{\boldsymbol{\theta}}_j \Vert \overset{p}{\to} 0$, which contradicts the condition $\Vert \sum_{j=1}^{t^0_{i_0}-1} \boldsymbol{K}' \boldsymbol{\theta}^0_j - \sum_{j=1}^{t^0_{i_0+1}-1} \boldsymbol{K}' \boldsymbol{\theta}^0_j \Vert > \nu > 0$ stated in Assumption \ref{as:5}. This implies $|\mathcal{A}_T| \geq m_0$ in probability and concludes the proof of the first part.

Turning to the second part, we define $\hat{T}_k = \lbrace \hat{t}_1, \hat{t}_2, \dots, \hat{t}_{k} \rbrace$. Then, it is enough to show that
\begin{eqnarray}
P \left( \lbrace d_H(\mathcal{A}_T, \mathcal{A}) > T\gamma_T, m_0 \leq |\mathcal{A}_T| \leq T \rbrace \right) \\
\notag = \sum\limits_{k=m_0}^{T} P \left( \lbrace d_H(\hat{T}_k, \mathcal{A}) > T\gamma_T \rbrace \right) P \left( |\mathcal{A}_T| = k \right) \to 0,
\end{eqnarray}
as $T \to \infty$. By Lemma \ref{lemma:5}, we have already shown that $P \left( d_H(\hat{T}_{m_0}, \mathcal{A}) > T\gamma_T \right) \to 0$ so that it suffices to show
\begin{equation}
\underset{k > m_0}{\max} \, P \left( d_H(\hat{T}_k, \mathcal{A}) > T\gamma_T \right) \to 0.
\end{equation}
Given $t^0_i$, we define
\begin{eqnarray}
\notag B_{T,k,i,1} &=& \lbrace \forall 1 \leq l \leq k, |\hat{t}_l - t^0_i| \geq T\gamma_T \text{ and } \hat{t}_l < t^0_i \rbrace \\
B_{T,k,i,2} &=& \lbrace \forall 1 \leq l \leq k, |\hat{t}_l - t^0_i| \geq T\gamma_T \text{ and } \hat{t}_l > t^0_i \rbrace \\
\notag B_{T,k,i,3} &=& \lbrace \exists 1 \leq l \leq k - 1, \text{ such that } |\hat{t}_l - t^0_i| \geq T\gamma_T, \\
\notag && \qquad |\hat{t}_{l+1} - t^0_i| \geq T\gamma_T \text{ and } \hat{t}_l < t^0_i < \hat{t}_{l+1} \rbrace.
\end{eqnarray}
Then, 
\begin{equation}
\underset{k > m_0}{\max} \, P \left( d_H(\hat{T}_k, \mathcal{A}) > T\gamma_T \right) = \underset{k > m_0}{\max} \, P \left( \bigcup\limits_{i=1}^{m_0} \bigcup\limits_{j=1}^{3} B_{T,k,i,j} \right).
\end{equation}
Using similar arguments as in the proof of Lemma \ref{lemma:5}, it can be shown that $\max_{k > m_0} \, P \left( \bigcup_{i=1}^{m_0} B_{T,k,i,j} \right) \to 0$ for $1 \leq j \leq 3$. This completes the proof of Theorem \ref{th:2}.
\end{proofof}
$\hfill \Box$


\begin{proofof}
We prove the first part of Theorem \ref{th:3} by showing that (a) $P \left( \widehat{\widehat{m}} < m_0 \right) \to 0$ and (b) $P \left( \widehat{\widehat{m}} > m_0 \right) \to 0$.

For part (a), Theorem \ref{th:2} implies that points $t_{Ti} \in \mathcal{A}_T$, $i = 1, \dots, m_0$ exist such that $\max_{1 \leq i \leq m_0} |\widehat{t}_{Ti} - t^0_i | \leq T\gamma_T$. This implies it is enough to show that if $m < m_0$, then
\begin{equation}
    IC(\widehat{\widehat{m}}, \widehat{\widehat{\boldsymbol{t}}}) \geq S_T(\widehat{t}_{T1}, \dots, \widehat{t}_{Tm_0}) + m_0 \omega_T,
\end{equation}
in probability. Let $R_T(m_0) = \lbrace (t_1, \dots, t_{m_0}): |t_i - t^0_i| \leq T\gamma_T, i = 1, \dots, m_0 \rbrace$. For any $\boldsymbol{t} \in R_T(m_0)$, we can write
\begin{eqnarray}
    \notag && S_T(t_1, \dots, t_{m_0}) \\
    \notag &=& \sum\limits_{i=1}^{t^0_1 - T\gamma_T - 1} \Vert Y_i - \bar{Z}_i \boldsymbol{K}' \widehat{\widehat{\boldsymbol{\theta}}}_1 \Vert^2 
    + \sum\limits_{j=2}^{m_0} \sum\limits_{i = t^0_{j-1} + T\gamma_T}^{t^0_j - T\gamma_T - 1} \Vert Y_i - \bar{Z}_i \sum\limits_{s=1}^{j} \boldsymbol{K}' \widehat{\widehat{\boldsymbol{\theta}}}_s \Vert^2 \\
    \notag &+& \sum\limits_{i = t^0_m + T\gamma_T}^{T} \Vert Y_i - \bar{Z}_i \sum\limits_{s=1}^{m_0+1} \boldsymbol{K}' \widehat{\widehat{\boldsymbol{\theta}}}_s \Vert^2 \\
    \notag &+& \sum\limits_{j=1}^{m_0} \sum\limits_{i = t^0_j - T\gamma_T}^{t^0_j - 1} \Vert Y_i - \bar{Z}_i \sum\limits_{s=1}^{j} \boldsymbol{K}' \widehat{\widehat{\boldsymbol{\theta}}}_s \Vert^2 + \sum\limits_{j=1}^{m_0} \sum\limits_{i = t^0_j}^{t^0_j + T\gamma_T - 1} \Vert Y_i - \bar{Z}_i \sum\limits_{s=1}^{j+1} \boldsymbol{K}' \widehat{\widehat{\boldsymbol{\theta}}}_s \Vert^2 \\
    &=& L_1 + L_2 + L_3 + L_4 + L_5. 
\end{eqnarray}
Since $\widehat{\widehat{\boldsymbol{\theta}}}_j$, $1 \leq j \leq m_0$ is the least squares estimator of $\boldsymbol{\theta}^0_j$ on the intervals $[1, t^0_1 - T\gamma_T - 1]$, $[t^0_{j-1} + T\gamma_T, t^0_j - T\gamma_T - 1]$, and $[t^0_m + T\gamma_T, T]$, it holds that
\begin{equation}
    L_1 + L_2 + L_3 \leq \left( \sum\limits_{i=1}^{t^0_1 - T\gamma_T - 1} + \sum\limits_{j=2}^{m_0} \sum\limits_{i = t^0_{j-1} + T\gamma_T}^{t^0_j - T\gamma_T - 1} + \sum\limits_{i = t^0_m + T\gamma_T}^{T} \right) \Vert u_i \Vert^2.
\end{equation}
Further, it can be shown that
\begin{equation}
    L_4 + L_5 \leq \left( \sum\limits_{j=1}^{m_0} \sum\limits_{i = t^0_j - T\gamma_T}^{t^0_j - 1} + \sum\limits_{j=1}^{m_0} \sum\limits_{i = t^0_j}^{t^0_j + T\gamma_T - 1} \right) \Vert u_i \Vert^2 + A_0 m_0 T \gamma_T,
\end{equation}
with probability approaching 1. Thus, with probability approaching one, 
\begin{equation}
    S_T(t_1, \dots, t_{m_0}) \leq \sum\limits_{i = 1}^{T} \Vert u_i \Vert^2 + A_0 m_0 T \gamma_T,
\end{equation}
holds uniformly for all $\boldsymbol{t} \in R_T(m_0)$. This implies that 
\begin{equation}
    S_T(\widehat{t}_{T1}, \dots, \widehat{t}_{Tm_0}) \leq \sum\limits_{i = 1}^{T} \Vert u_i \Vert^2 + A_0 m_0 T \gamma_T,
    \label{eq:SSE_lower}
\end{equation}
holds with probability approaching one. However, it can be shown using similar arguments as in Lemma A.4 in \cite{ChanYauZhang2014} that if $m < m_0$, then
\begin{equation}
    S_T(\widehat{\widehat{t}}_1, \dots, \widehat{\widehat{t}}_m) \geq \sum\limits_{i = 1}^{T} \Vert u_i \Vert^2 + \nu ( \underset{1 \leq i \leq m_0}{\min} \vert t^0_i - t^0_{i-1} \vert ),  
    \label{eq:SSE_upper}
\end{equation}
with probability approaching one. Combining results \eqref{eq:SSE_lower} and \eqref{eq:SSE_upper}, we obtain with probability approaching one that
\begin{eqnarray}
    IC(\widehat{\widehat{m}}, \widehat{\widehat{\boldsymbol{t}}}) &=& S_T(\widehat{\widehat{t}}_1, \dots, \widehat{\widehat{t}}_m) + m \omega_T \\
    \notag &\geq& \sum\limits_{i = 1}^{T} \Vert u_i \Vert^2 + \nu ( \underset{1 \leq i \leq m_0}{\min} \vert t^0_i - t^0_{i-1} \vert ) + m \omega_T \\
    \notag &\geq& S_T(\widehat{t}_{T1}, \dots, \widehat{t}_{Tm_0}) + m_0 \omega_T + \nu ( \underset{1 \leq i \leq m_0}{\min} \vert t^0_i - t^0_{i-1} \vert ) - A_0 m_0 T \gamma_T - (m_0 - m) \omega_T \\
    \notag &\geq& S_T(\widehat{t}_{T1}, \dots, \widehat{t}_{Tm_0}) + m_0 \omega_T,
\end{eqnarray}
where the last inequality follows from the conditions $\omega_T/\min_{1 \leq i \leq m_0} \vert t^0_i - t^0_{i-1} \vert \to 0$ and $\lim_{T \to \infty} T\gamma_T / \omega_T \leq 1$. This result implies that $P \left( \widehat{\widehat{m}} < m_0 \right) \to 0$ for $T \to \infty$ and concludes the proof of part (a).

Next, we turn to part (b). Here, it is enough to show that if $m > m_0$, then $IC(m, \widehat{\widehat{t}}_{1}, \dots, \widehat{\widehat{t}}_m) > IC(m_0, \widehat{\widehat{t}}_{1}, \dots, \widehat{\widehat{t}}_{m_0})$. We note that, when $m > m_0$, it holds that
\begin{eqnarray}
    S_T(\widehat{t}_{T1}, \dots, \widehat{t}_{Tm_0}) &\geq& S_T(\widehat{\widehat{t}}_{1}, \dots, \widehat{\widehat{t}}_{m_0}) \geq S_T(\widehat{\widehat{t}}_{1}, \dots, \widehat{\widehat{t}}_m)\\
    &\geq& S_T(\widehat{\widehat{t}}_{1}, \dots, \widehat{\widehat{t}}_m, t^0_1, \dots, t^0_{m_0}).
    \label{eq:SSE_ineq}
\end{eqnarray}
It can be shown that
\begin{equation}
    S_T(\widehat{\widehat{t}}_{1}, \dots, \widehat{\widehat{t}}_m, t^0_1, \dots, t^0_{m_0}) \geq \sum\limits_{i = 1}^{T} \Vert u_i \Vert^2 - (m + m_0) T \gamma_T.
    \label{eq:SSE_more_upper}
\end{equation}
From Equation~\ref{eq:SSE_ineq}, it follows that 
\begin{eqnarray}
    S_T(\widehat{t}_{T1}, \dots, \widehat{t}_{Tm_0}) - S_T(\widehat{\widehat{t}}_1, \dots, \widehat{\widehat{t}}_m) \geq S_T(\widehat{\widehat{t}}_1, \dots, \widehat{\widehat{t}}_{m_0}) - S_T(\widehat{\widehat{t}}_1, \dots, \widehat{\widehat{t}}_m) \geq 0,
\end{eqnarray}
and, since we have established a upper bound for $S_T(\widehat{t}_{T1}, \dots, \widehat{t}_{Tm_0})$ in Equation~\ref{eq:SSE_more_upper} and a lower bound for $S_T(\widehat{\widehat{t}}_{1}, \dots, \widehat{\widehat{t}}_m)$ in Equation~\ref{eq:SSE_lower}, it holds with probability approaching one that
\begin{equation}
    S_T(\widehat{\widehat{t}}_1, \dots, \widehat{\widehat{t}}_{m_0}) - S_T(\widehat{\widehat{t}}_1, \dots, \widehat{\widehat{t}}_m) \geq (m + m_0 (1 + A_0)) T \gamma_T.
\end{equation}
This yields
\begin{equation}
    IC(m, \widehat{\widehat{t}}_1, \dots, \widehat{\widehat{t}}_m) - IC(m_0, \widehat{\widehat{t}}_{1}, \dots, \widehat{\widehat{t}}_{m_0}) \geq (m - m_0) \omega_T - (m + m_0 (1 + A_0)) T \gamma_T > 0,
\end{equation}
in probability because of the condition $\lim_{T \to \infty} T \gamma_T / \omega_T = 0$. This implies $P \left( \widehat{\widehat{m}} > m_0 \right) \to 0$ for $T \to \infty$ and concludes the proof of part (b).

For the second property, if there exists one $1 \leq i \leq m_0$ such that $|\widehat{\widehat{t}}_i - t^0_i| \geq 2 A_0 m_0 T \gamma_T/\nu = B T \gamma_T$, then there must exist a $t^0_l$ such that $|\widehat{\widehat{t}}_i - t^0_l| \geq B T \gamma_T$ for all $1 \leq i \leq m_0$ and it can be shown that
\begin{equation}
    S_T(\widehat{\widehat{t}}_1, \dots, \widehat{\widehat{t}}_{m_0}) \geq \sum\limits_{i = 1}^{T} \Vert u_i \Vert^2 + 2 A_0 m_0 T \gamma_T.
\end{equation}
However, this contradicts Equation~\eqref{eq:SSE_lower} and thus $P \left( \max_{1 \leq i \leq m_0} \vert \widehat{\widehat{t}}_i - t^0_i \vert \leq B T \gamma_T \right) \to 1$.
\end{proofof}
$\hfill \Box$


\begin{proofof}
Again, we can decompose the proof of the first property into two parts, (a) $P ( |\mathcal{A}^*_T| < m_0 ) \to 0$ and (b) $P ( |\mathcal{A}^*_T| > m_0 ) \to 0$. Using similar arguments as in the proof of Theorem \ref{th:3}, it follows that $P ( |\mathcal{A}^*_T| < m_0 ) \to 0$ for $T \to \infty$. Turning to part (b), Suppose that $\mathcal{A}^*_T = m > m_0$, then 
\begin{equation}
    S (\mathcal{A}^*_T) = \underset{(t_1, \dots, t_m) \subseteq \mathcal{A}_T}{\min} S_T(t_1, \dots, t_m),
\end{equation}
which implies
\begin{eqnarray}
    S_T(\widehat{t}^*_1, \dots, \widehat{t}^*_m) \leq S_T(\widehat{t}_1, \dots, \widehat{t}_m) \leq S_T(\widehat{t}_1, \dots, \widehat{t}_{m_0}) \leq \sum\limits_{i = 1}^{T} \Vert u_i \Vert^2 + m_0 \mathcal{V} T \gamma_T,
\end{eqnarray}
with probability approaching one. However, we also have
\begin{equation}
    S_T(\widehat{t}^*_1, \dots, \widehat{t}^*_m) \geq S_T(\widehat{t}^*_1, \dots, \widehat{t}^*_m, t^0_1, \dots, t^0_{m_0}) \geq \sum\limits_{i = 1}^{T} \Vert u_i \Vert^2 - (m + m_0) T \gamma_T,
\end{equation}
with probability approaching one. Using the previous inequalities as in Theorem \ref{th:3}, it follows that $P ( |\mathcal{A}^*_T| > m_0 ) \to 0$ which concludes the proof of the first property. The proof of the second property follows as in Theorem \ref{th:3}.
\end{proofof}
$\hfill \Box$



\end{document}